\documentclass[12pt]{article}
\usepackage{amsfonts,amssymb,amsmath,mathtools,graphics,slashed}
\usepackage{hyperref,hep,graphicx,subfig}
\usepackage{rotating}
\usepackage{xcolor}
\usepackage{verbatim}
\usepackage[final]{showlabels}

\definecolor{greenish}{RGB}{0,190,0}

\definecolor{yellowish}{RGB}{190,190,0}

\definecolor{bluish}{RGB}{0,0,190}

\newcommand{\nn}{\notag \\}

\makeatletter

\@addtoreset{equation}{section}
\makeatother

\begin{document}

\begin{titlepage}

\vfill


\vfill

\begin{center}
   \baselineskip=16pt
   {\Large\bf Higgs/Amplitude Mode Dynamics From Holography}
  \vskip 1.5cm
  \vskip 1.5cm
      Aristomenis Donos$^1$ and Christiana Pantelidou$^2$\\
   \vskip .6cm
      \begin{small}
      \textit{$^1$ Centre for Particle Theory and Department of Mathematical Sciences,\\ Durham University,
       Durham, DH1 3LE, U.K.}\\
        \textit{$^2$ School of Mathematics and Statistics,\\ University College Dublin, Belfield, Dublin 4, Ireland}
        \end{small}\\   
         
\end{center}

\vfill

\begin{center}
\textbf{Abstract}
\end{center}
\begin{quote}
Second order phase transitions are universally driven by an order parameter which becomes trivial at the critical point. At the same time, collective excitations which involve the amplitude of the order parameter develop a gap which smoothly closes to zero at criticality. We develop analytical techniques to study this ``Higgs" mode in holographic systems which undergo a continuous phase transition at finite temperature and chemical potential. This allows us to study the linear response of the system at energy scales of the order of the gap. We express the Green's functions of scalar operators in terms of thermodynamic quantities and a single transport coefficient which we fix in terms of black hole horizon data.
\end{quote}

\vfill

\end{titlepage}

\setcounter{equation}{0}

\section{Introduction}
Holography provides a powerful framework to study strongly coupled quantum field theories \cite{Aharony:1999ti} at finite temperature and chemical potential \cite{Hartnoll:2016apf}. In particular, in the large $N$ limit, it allows us to perform exact computations which are impossible to carry out with standard quantum field theory techniques.

Second order phase transitions are driven by an order parameter which becomes non trivial in the broken phase. Depending on the nature of the system, the phase transition can be accompanied by the breaking of a global symmetry. In such a case, new gapless modes appear in the spectrum of the system in the form of Goldstone modes. The universal feature is that the phase transition is always accompanied by the emergence of a gapped mode which is exactly gapless at the critical point, the amplitude/Higgs mode (from now on we will call it the Higgs mode). At infinite wavelengths, the corresponding mode\footnote{Strictly speaking, the mode we are referring to is an amplitude mode for the order parameter of the system.} is homogeneous and it is related to fluctuating degrees of freedom which are intimately connected to the amplitude of the order parameter and which decouple from the entropy and charge density.

The above highlights that the Higgs mode dominates critical phenomena as its decay rate is parametrically small close to the critical point \cite{2015ARCMP...6..269P, RevModPhys.49.435}. Despite its fundamental importance, the Higgs mode was observed in condensed matter systems only very recently. For the case of superfluids see e.g. \cite{Sherman:2015wv,Endres:2012vf}. In view of these results, it is very important to obtain theoretical predictions on potentially universal properties of the critical dynamics. In this paper, we will study aspects of the dynamics close to a second order phase transition by exploiting the holographic principle.

In holography, continuous phase transitions can be realised through black hole instabilities. With fixed deformation parameters and chemical potential, a unique black hole solution dominates the phase diagram at high temperatures. Lowering the Hawking temperature, certain bulk perturbations can become unstable below a critical temperature $T_c$. Following the instability leads to new branches of black hole solutions corresponding to the broken phase. In the context of applied holography, some of the most well studied examples are the holographic superfluids \cite{Hartnoll:2008vx,Gubser:2008px,Gubser:2008zu} and phases which spontaneously break translations \cite{Donos:2011bh, Donos:2011qt , Donos:2013gda}.

It is natural to ask how much we can learn about the Higgs mode by considering holographic systems near their critical points. The Higgs mode has been identified in several holographic studies in the past by  using numerical techniques \cite{Amado:2009ts,Bhaseen:2012gg} or analytically when full solutions in the bulk can be obtained \cite{Herzog:2010vz}. Given the power of holography in obtaining exact results, the ultimate goal would be to construct an enlarged theory of hydrodynamics which incorporates this universal nearly gapless mode in its description.

In this paper we will make a significant step in this direction by examining the dynamics of holographic systems close to their phase transitions. More specifically, we will obtain analytic expressions for the low frequency retarded Green's functions of scalar operators. As we show in section \ref{sec:greens}, these are entirely determined by the thermodynamic properties of the system and a single transport coefficient which is fixed by the black hole horizon data. The linear response of the system is dominated by a single pole which corresponds to the Higgs mode which decays slowly near the critical point. Our main tool will be the techniques which have been recently exploited in \cite{Donos:2021pkk} to obtain a hydrodynamic description of holographic superfluids at zero charge density. The main contribution was to write the transport coefficients of \cite{KHALATNIKOV198270} in terms of black hole horizon data.

In that construction, the basic ingredient is the symplectic current density of the bulk theory \cite{Crnkovic:1986ex}. This can be considered for any pair of perturbative solutions in the bulk. Being a generalisation of Liouville's theorem for classical mechanics, the symplectic current is divergence free. This feature will be very important to us since it allows us to extract useful information for an unknown perturbative solution when it is parametrically close to a known one.

In our context, the role of the known solution will be played by the static solutions we can construct by varying the thermodynamic parameters of the thermal state close to the critical point. At exactly the critical point, this variation reduces to the static bulk perturbative solution that emerges at the critical point and which leads to the broken phase black holes. This, is also the limit of the time dependent Higgs mode we are after when the system is exactly at the transition point. This suggests that the bulk dual of the Higgs mode is parametrically close to the solution generated by varying the thermodynamic parameters and will therefore be the unknown solution in the symplectic current.

The paper is organised in five sections as follows. In Section \ref{sec:setup} we introduce the  holographic framework in which we plan to study phase transitions driven by a scalar order parameter. In this case, the phase transition is not accompanied by the breaking of a continuous symmetry. In Section \ref{sec:Higgs_mode} we analyse the time dependent perturbation which captures the Higgs mode and we use our technique involving the symplectic current to extract its gap. In Section \ref{sec:superfluids} we generalise our holographic model to describe the breaking of a global symmetry on the boundary at finite chemical potential. This is particularly interest since, as we will see, the lack of particle-hole symmetry mixes the Goldstone mode. In Section \ref{sec:greens} we compute the Green's functions of the scalar operators in our holographic theory highlighting their domination by the pole relevant to the Higgs mode. Finally, in Section \ref{sec:numerics} we perform numerical checks of our analytic results for the gap and the Green's functions we compute in Section \ref{sec:greens}. We conclude the paper with a summary of our main results and outlook in Section \ref{sec:discussion}.

\section{Setup}\label{sec:setup}
We will consider a bulk theory which can model the phase transition of a conformal field theory at finite chemical potential and which has been deformed by a scalar operator $\mathcal{O}_\phi$ of dimension $\Delta_\phi$ with bulk dual $\phi$. In addition to that, our bulk theory will contain the metric $g_{\mu\nu}$,  gauge field $A_\mu$ and scalar field $\rho$ which will spontaneously give a VEV to the field theory dual operator $\mathcal{O}_\rho$.

Without loss of generality, we will consider the bulk action,
\begin{align}\label{eq:bulk_action}
S_{b}=\int d^4x\,\sqrt{-g}\,\left( R-V-\frac{1}{2}\left(\partial\phi \right)^2-\frac{1}{2}\left(\partial\rho \right)^2-\frac{Z}{4}F^2\right)\,,
\end{align}
where $F=dA$ is the field strength of $A_\mu$. In general, the gauge coupling $Z$ and potential $V$ can depend on the scalar fields $\phi$ and $\rho$. In order for the bulk equations of motion to admit solutions which asymptote to unit radius $AdS_4$, we will impose that for small values of the scalar fields,
\begin{align}
V&\approx 6-\frac{1}{2}m_\phi^2\,\phi^2-\frac{1}{2}m_\rho^2\,\rho^2+\cdots\,,\notag\\
Z&\approx 1+\zeta_\phi\,\phi^2+\zeta_\rho\,\rho^2+\cdots\,.
\end{align}
Moreover, without any loss of generality, for small values of $\rho$ we can assume that,
\begin{align}
V&\approx v(\phi)-\frac{1}{2}m^2(\phi)\,\rho^2-\frac{1}{4!}\lambda(\phi)\,\rho^4+\cdots\,,\notag\\
Z&\approx \zeta_0(\phi)+\zeta_1(\phi)\,\rho^2+\cdots\,.
\end{align}

The background black holes that can capture the different phases of our theory are described by the background geometries with,
\begin{align}\label{eq:background}
ds^{2}&=-U(r)\,dt^{2}+\frac{dr^{2}}{U(r)}+e^{2g(r)}\,\left(dx^{2}+dy^{2} \right)\,,\nn
\phi&=\phi(r),\qquad \rho=\rho(r), \qquad A=a(r)\,dt\,.
\end{align}
The scalar $\rho$ is identically zero in the normal phase, above the critical temperature $T_c$.

We will choose coordinates so that the event horizon with Hawking temperature $T$ is at $r=0$. Near the horizon, regularity dictates the Taylor expansions,
\begin{align}\label{eq:nh_bcs}
&U(r)\approx\,4\pi T\,r+\mathcal{O}(r^{2}),\qquad g(r)\approx g^{(0)}+\mathcal{O}(r)\,,\nn
&\phi(r)\approx \phi^{(0)}+\mathcal{O}(r),\qquad \rho(r)\approx \rho^{(0)}+\mathcal{O}(r)\,,\nn
&a(r)\approx a^{(0)}\,r+\cdots\,.
\end{align}
At the other end of the geometry, where $r\to\infty$, we want to impose $AdS_4$ asymptotics with unit radius,
\begin{align}\label{eq:uv_bcs}
&U(r)\approx (r+R)^{2}+\mathcal{O}(r^{0}),\qquad g(r)\approx \ln(r+R)+\mathcal{O}(r^{-1})\,,\nn
&\phi(r)\approx \phi_{(s)}\,(r+R)^{\Delta_{\phi}-3}+\cdots +\phi_{(v)}\,(r+R)^{-\Delta_{\phi}}+\cdots\,,\nn
&\rho(r)\approx \rho_{(s)}\,(r+R)^{\Delta_{\rho}-3}+\cdots  \rho_{(v)}\,(r+R)^{-\Delta_{\rho}}+\cdots\,,\nn
&a(r)\approx \mu-\varrho\,(r+R)^{-1}+\cdots\,.
\end{align}
The conformal dimensions $\Delta_\phi$ and $\Delta_\rho$ of the dual operators $\mathcal{O}_\phi$ and $\mathcal{O}_\rho$ are determined in terms of the masses according to $\Delta_\phi (\Delta_\phi-3)=m_\phi^2$ and $\Delta_\rho (\Delta_\rho-3)=m_\rho^2$. The constants of integration $\phi_{(s)}$ and $\rho_{(s)}$ correspond to sources for the boundary operator $\mathcal{O}_\phi$ and the ``modulus" of the operator $\mathcal{O}_\rho$. We will be mostly setting the latter to zero except for section \ref{sec:greens} where will turn it on perturbatively in order to extract a few of the retarded Green's functions of the system.

At this point, it is useful to note that the horizon charge density $\varrho^{(0)}$ and entropy density $s$ satisfy,
\begin{align}\label{eq:cs_horizon}
\varrho^{(0)}=Z^{(0)}\,e^{2\,g^{(0)}}\,a^{(0)},\qquad s=4\pi\,e^{2\,g^{(0)}}\,,
\end{align}
where $Z^{(0)}$ denotes the value of the coupling function $Z$ evaluated on the horizon values of the background scalars $\phi$ and $\rho$. For normal fluids, the horizon charge density $\rho^{(0)}$ coincides with the field theory charge density $\varrho=\varrho_{(v)}$ which can be extracted from the asymptotics \eqref{eq:uv_bcs}. For the superfluid phase that we will consider in section \ref{sec:superfluids}, this is no longer true. In contrast, the entropy density can always be computed as a horizon quantity from \eqref{eq:cs_horizon}.

\section{The Higgs Mode from Holography}\label{sec:Higgs_mode}

In this section we will develop our tools that will lead to the main results of our paper. The time dependent perturbation capturing the Higgs mode we are after will be constructed in an expansion of the thermodynamic and deformation parameters around the transition point. In section \ref{sec:static_expansions} we will discuss the static perturbations we can obtain by varying the thermodynamic and deformation parameters of the black hole solutions which are dual to the thermal states of our system. These will prove particularly useful in constructing the next to leading order terms in the expansion of the time dependent perturbation in section \ref{sec:t_pert}. Finally, in section \ref{sec:sympl_current} we will make use of the symplectic current to fix the frequency of the quasinormal mode as a function of our thermodynamics and black hole horizon data.

\subsection{Expansions Around The Critical Point}\label{sec:static_expansions}

The gapped mode that we will study in the following sections corresponds to a time dependent perturbation around the background geometry \eqref{eq:background}.  We will be interested in studying this perturbation very close to the critical point $(T_c(\mu,\phi_{(s)}),\mu,\phi_{(s)})$.  For this reason, we will introduce a parameter $\varepsilon$ which parametrises a curve in the $(T,\mu,\phi_{(s)})$ plane that originates from the point $(T_c(\mu,\phi_{(s)}),\mu,\phi_{(s)})$. As we vary the parameter $\varepsilon$ we move in the space of thermodynamic parameters according to $(T_c(\mu,\phi_{(s)})+\delta T(\varepsilon),\mu+\delta\mu(\varepsilon),\phi_{(s)}+\delta\phi_{(s)}(\varepsilon))$. To make our notation clear we consider the curve for small values of the expanision parameter with,
\begin{align}
\delta T(\varepsilon)\approx \delta T\,\frac{\varepsilon^2}{2},\qquad \delta\mu(\varepsilon)\approx \delta\mu\,\frac{\varepsilon^2}{2},\qquad \delta\phi_{(s)}(\varepsilon)\approx\delta\phi_{(s)}\frac{\varepsilon^2}{2}\,.
\end{align}
For a general function $\Phi$ of the thermodynamic parameters, along the curve we can write the variation $\delta \Phi=\frac{\varepsilon^2}{2}\,\delta\Phi_{(2)}+\cdots$ with,
\begin{align}
\delta\Phi_{(2)}=\delta T\,\partial_T\Phi+\delta\mu\,\partial_\mu\Phi+\delta\phi_{(s)}\,\partial_{\phi_{(s)}}\Phi\,.
\end{align}

Right at the critical point, there exists a static perturbative solution $\delta\rho_{\ast(0)}$ to the equation of motion for the scalar $\rho$ with source free boundary conditions, just like in  \eqref{eq:uv_bcs} for the non-linear problem. Our goal is to track this perturbative mode right below the critical temperature where it acquires a gap $\delta\omega_{g}$.  As we will see later more explicitly  the gap is itself an expandable function of $\varepsilon$ and we will have $\delta\omega_{g}\propto \varepsilon^2$.

In studying our gapped mode, we will find useful to consider the expansion of the background geometry itself in $\varepsilon$. Constructing this perturbative expansion is a rather non-trivial task. However, here we will only need the fact that it exists and that it is analytic in $\varepsilon$,
\begin{align}\label{eq:broken_exp}
\rho&=\varepsilon_{\ast}\,\delta\rho_{\ast(0)}+\frac{\varepsilon_{\ast}^3}{3!}\,\delta\rho_{\ast(2)}+\cdots\,,\nn
U&=U_c+\frac{\varepsilon_{\ast}^2}{2}\,\delta U_{\ast(2)}+\frac{\varepsilon_{\ast}^4}{4!}\,\delta U_{\ast(4)}+\cdots,\quad g=g_c+\frac{\varepsilon_{\ast}^2}{2}\,\delta g_{\ast(2)}+\frac{\varepsilon_{\ast}^4}{4!}\,\delta g_{\ast(4)}+\cdots\,,\nn
\phi&=\phi_c+\frac{\varepsilon_{\ast}^2}{2}\,\delta \phi_{\ast(2)}+\frac{\varepsilon_{\ast}^4}{4!}\,\delta \phi_{\ast(4)}+\cdots\,,\quad a=a_c+\frac{\varepsilon_{\ast}^2}{2}\,\delta a_{\ast(2)}+\frac{\varepsilon_{\ast}^4}{4!}\,\delta a_{\ast(4)}+\cdots\,.
\end{align}
The leading terms in the above expansion represent the black hole solution \eqref{eq:background} at the critical point.  For the curve of variation in the broken phase we define the variation parameter $\varepsilon_\ast$ according to,
\begin{align}\label{eq:var_broken}
\delta T(\varepsilon_\ast)\approx \delta T_{\ast(2)}\,\frac{\varepsilon_\ast^2}{2}+\cdots,\quad \delta\mu(\varepsilon_\ast)\approx \delta\mu_{\ast(2)}\,\frac{\varepsilon_\ast^2}{2}+\cdots,\quad
\delta\phi_{(s)}(\varepsilon_\ast)\approx\delta\phi_{(s)\ast (2)}\frac{\varepsilon_\ast^2}{2}\,.
\end{align}

Apart from the broken phase black holes,  it is useful to consider the expansion of the normal phase black hole solutions below $T_c$, 
\begin{align}\label{eq:normal_exp}
\rho&=0\,,\nn
U&=U_c+\frac{\varepsilon_{\#}^2}{2}\,\delta U_{\#(2)}+\frac{\varepsilon_{\#}^4}{4!}\,\delta U_{\#(4)}+\cdots,\quad g=g_c+\frac{\varepsilon_{\#}^2}{2}\,\delta g_{\#(2)}+\frac{\varepsilon_{\#}^4}{4!}\,\delta g_{\#(4)}+\cdots\,,\nn
\phi&=\phi_c+\frac{\varepsilon_{\#}^2}{2}\,\delta \phi_{\#(2)}+\frac{\varepsilon_{\#}^4}{4!}\,\delta \phi_{\#(4)}+\cdots\,,\quad a=a_c+\frac{\varepsilon_{\#}^2}{2}\,\delta a_{\#(2)}+\frac{\varepsilon_{\#}^4}{4!}\,\delta a_{\#(4)}+\cdots\,.
\end{align}
Similarly to the case of the broken phase, we will define the expansion parameter $\varepsilon_\#$ according to,
\begin{align}\label{eq:var_normal}
\delta T(\varepsilon_\#)\approx \delta T_{\#(2)}\,\frac{\varepsilon_\#^2}{2}+\cdots,\quad \delta\mu(\varepsilon_\#)\approx \delta\mu_{\#(2)}\,\frac{\varepsilon_\#^2}{2}+\cdots,\quad
\delta\phi_{(s)}(\varepsilon_\#)\approx\delta\phi_{(s)\# (2)}\frac{\varepsilon_\#^2}{2}\,.
\end{align}

It is also useful to write thermodynamic quantities in terms of the perturbative expansions of the horizon data.  In particular, we can simply expand the charge and the entropy densities of equation \eqref{eq:cs_horizon} in $\varepsilon$ to obtain,
\begin{align}
\delta T_\ast&=\frac{\varepsilon_\ast^2}{2}\,\delta T_{\ast(2)}+\cdots,\quad \delta\varrho_\ast=\frac{\varepsilon_{\ast}^2}{2}e^{2g_c^{(0)}}\,\left( \delta a^{(0)}_{\ast(2)}+2\,\delta g_{\ast(2)}^{(0)}\,a^{(0)}\right)+\cdots\,,\notag\\
\delta s_\ast&=\varepsilon_{\ast}^24\pi\,e^{2g_c^{(0)}}\,\delta g_{\ast(2)}^{(0)}+\cdots\,,\nn
\delta T_\#&=\frac{\varepsilon_{\#}^2}{2}\,\delta T_{\#(2)}+\cdots,\quad \delta\varrho_\#=\frac{\varepsilon_{\#}^2}{2}e^{2g_c^{(0)}}\,\left( \delta a^{(0)}_{\#(2)}+2\,\delta g_{\#(2)}^{(0)}\,a^{(0)}\right)+\cdots\,,\notag\\
\delta s_\#=&\varepsilon_{\#}^24\pi\,e^{2g_c^{(0)}}\,\delta g_{\#(2)}^{(0)}+\cdots\,.
\end{align}
The above horizon constants come from e.g. the near horizon expansions
\begin{align}
\delta U_{\ast(n)}(r)&=4\pi\,\delta T_{\ast(n)}\,r+\cdots,\quad \delta g_{\ast(n)}(r)=\delta g_{\ast(n)}^{(0)}+\cdots\,,\nn
\delta a_{\ast(n)}(r)&=\delta a_{\ast(n)}^{(0)}\,r+\cdots,\quad \delta\phi_{\ast (n)}=\delta\phi_{\ast (n)}^{(0)}+\cdots\,,
\end{align}
and similarly for the normal phase $\varepsilon$ expansion. Close to the conformal boundary we will have the asymptotic expansions,
\begin{align}\label{eq:uv_bcs_pert}
&\delta U_{\ast(n)}(r)\approx \mathcal{O}(r^{0}),\qquad \delta g_{\ast(n)}(r)\approx \mathcal{O}(r^{-1})\,,\nn
&\delta\phi_{\ast(n)}(r)\approx \delta\phi_{(s)\ast(n)}\,(r+R)^{\Delta_{\phi}-3}+\cdots +\delta\phi_{(v)\ast(n)}\,(r+R)^{-\Delta_{\phi}}+\cdots\,,\nn
&\delta\rho_{\ast(n)}(r)\approx \delta\rho_{(s)\ast(n)}\,(r+R)^{\Delta_{\rho}-3}+\cdots  \delta\rho_{(v)\ast(n)}\,(r+R)^{-\Delta_{\rho}}+\cdots\,,\nn
&\delta a_{\ast(n)}(r)\approx \delta\mu_{\ast(n)}-\delta\varrho_{\ast(n)}\,(r+R)^{-1}+\cdots\,.
\end{align}
We will mostly keep $\delta\rho_{(s)\ast(n)}=0$, except in Section \ref{sec:greens} where we will study the retarded Green's functions of our scalar operators. Note that similar expansions hold for the functions appearing in the expansion around the normal phase black holes.

For the VEVs of the scalar operators we can write the expansions,
\begin{align}\label{eq:vev_vars}
&\delta \langle\mathcal{O}_\rho\rangle_\ast=\varepsilon_\ast\,\delta \langle\mathcal{O}_\rho\rangle_{\ast(0)}+\frac{\varepsilon_\ast^3}{3!}\,\delta \langle\mathcal{O}_\rho\rangle_{\ast(2)}+\cdots,\quad \delta \langle\mathcal{O}_\phi\rangle_\ast=\frac{\varepsilon_\ast^2}{2}\,\delta \langle\mathcal{O}_\phi\rangle_{\ast(2)}+\cdots\,,\notag\\
&\delta \langle\mathcal{O}_\phi\rangle_\#=\frac{\varepsilon_\#^2}{2}\,\delta \langle\mathcal{O}_\phi\rangle_{\#(2)}+\cdots\,.
\end{align}

\subsection{Time Dependent Perturbation}\label{sec:t_pert}

In this section we will consider perturbative solutions of the equations of motion around the backgrounds \eqref{eq:background}.  Since our background geometries possess spacetime translational invariance, it makes sense to consider the Fourier decomposition of our perturbations. Since we are only interested in extracting the gap, we will only consider Fourier modes with zero wavenumber according to,
\begin{align}\label{eq:qnm_ansatz}
\delta \mathcal{F}(t,\mathbf{x};r)=e^{-i\omega (t+S(r))}\,\delta f(r)\,,
\end{align}
where $\mathcal{F}$ repsesents perturbations of the scalars as well as the metric and gauge field components. 
The function $S(r)$ is chosen so that it drops faster that $\mathcal{O}(1/r^3)$ close to the conformal boundary and it therefore doesn't interfere with holographic renormalisation. However, close to the horizon, it is chosen so that it approaches $S(r)\to \frac{1}{4\pi T}\ln r$ and the combination $t+S(r)$ is regular and ingoing. 

For the system we are interested in, it is consistent to set the component perturbations $\delta g_{t i}$,  $\delta g_{r i}$, $\delta g_{xy}$ and $\delta a_i$ to zero.  Moreover, due to the unbroken rotational symmetry in the $x$-$y$ plane we can set $\delta g_{xx}=\delta g_{yy}$. Imposing regular ingoing boundary conditions close to the horizon, leads to the near horizon expansions,
\begin{align}\label{eq:gen_exp}
\delta g_{tt}(r)&= 4\pi T\,r\, \delta g_{tt}^{(0)}+\cdots\,,\quad
\delta g_{rr}(r)=\frac{\delta g_{rr}^{(0)}}{4\pi T\,r}+\cdots\, \,,\nn
\delta g_{ii}(r)&=\delta g^{(0)}+r\,\delta g^{(1)}+\cdots\,,\quad\quad
\delta g_{tr}(r)=\delta g_{tr}^{(0)}+\cdots\,,\nn
\delta a_{t}(r)&=\delta a_{t}^{(0)}+\delta a_{t}^{(1)}\,r+\cdots\,,\quad
\delta a_{r}(r)=\frac{1}{4\pi T\,r} \delta a_{r}^{(0)}+\delta a_{r}^{(1)}+\cdots\,,\nn
\delta \rho(r)&=\delta \rho^{(0)}+\cdots,\quad \delta \phi(r)=\delta \phi^{(0)}+\cdots\,,
\end{align}
which are compatible with the equations of motion. In order to achieve regularity, the above conditions need to be supplemented by,
\begin{align}\label{eq:nh_reg}
\delta g_{tt}^{(0)}+\delta g_{rr}^{(0)}=2\,\delta g_{rt}^{(0)}\,,\quad
\delta a_{r}^{(0)}=\delta a_{t}^{(0)}\,.
\end{align}

The time dependent perturbation which is capturing the Higgs mode can be expanded according to\footnote{Notice that we don't discuss $\delta \tilde{g}_{tr}$ in the expansion \eqref{eq:timed_pert} as this can be changed by a coordinate transformation of the form $t\rightarrow t+\delta f(r)$ for a perturbatively small function $\delta f(r)$ and can be freely chosen as long as it satisfies the boundary conditions \eqref{eq:gen_exp} and \eqref{eq:nh_reg}.},
\begin{align}\label{eq:timed_pert}
\omega&=\varepsilon_\ast\,\omega_{[1]}+\varepsilon_\ast^2\,\omega_{[2]}+\cdots\,,\nn
\delta \tilde{g}_{tt}&=-\varepsilon_\ast\,\delta\tilde{U}_{(2)}+\cdots,\quad \delta \tilde{g}_{rr}=-\varepsilon_\ast\,\frac{\delta \tilde{U}_{(2)}}{U_c^2}+\cdots\,,\nn
\delta\tilde{g}_{ii}&=2\,\,e^{2g_c}\,\left(\varepsilon_\ast\,\delta \tilde{g}_{{2}}+\cdots\right),\quad \delta\tilde{a}_t=\varepsilon_\ast\,\delta \tilde{a}_{(2)}+\cdots\,,\nn
\delta\tilde{\phi}&=\varepsilon_\ast\,\delta\tilde{\phi}_{(2)}+\cdots,\quad\delta\tilde{\rho}=\delta\rho_{\ast(0)}+\frac{\varepsilon_\ast^2}{2}\delta\tilde{\rho}_{(2)}+\cdots\,.
\end{align}
The above expansion makes clear that our perturbation has to go to the zero mode $\delta\rho_{\ast(0)}$ at the critical point with $\varepsilon_\ast=0$. Notice that the choice of $\delta \tilde{g}_{rr}$ in the ansatz \eqref{eq:timed_pert} fixes the choice of the radial coordinate at order $\mathcal{O}(\varepsilon_\ast)$. The functions in equation \eqref{eq:timed_pert} satisfy the boundary conditions \eqref{eq:gen_exp} and \eqref{eq:nh_reg}. For $\varepsilon_\ast=0$ the mode becomes exactly the static mode which drives the transition. This is precisely the way the above expansion is singling out the specific mode we wish to study from the rest of the tower of the quasi-normal modes that we expect to exist in the lower half-plane.

As we will see later, the leading term in the frequency expansion $\omega_{[1]}$ is equal to zero.  An  important point in understanding the first non-trivial $\varepsilon_\ast$ correction in our perturbation \eqref{eq:timed_pert} is that the equations that the functions $\delta\tilde{U}_{(2)}$, $\delta\tilde{g}_{(2)}$, $\delta\tilde{a}_{(2)}$, $\delta\tilde{\phi}_{(2)}$ and $\delta\tilde{\rho}_{(2)}$ satisfy are a superset of the equations saitisfied by the functions $\delta U_{\ast(2)}$, $\delta g_{\ast(2)}$, $\delta a_{\ast(2)}$, $\delta \phi_{\ast(2)}$ and $\delta \rho_{\ast(2)}$. The two extra equations are constraints and impose charge and energy conservation from the field theory point of view. As such these constraints can be imposed on any constant $r$ surface. However, as we will see in section \ref{sec:superfluids}, this is no longer true for the equation of motion of the 1-form field due to the fact that it becomes massive in the bulk. This is directly related to the spontaneous symmetry breaking and the involvement of the field theory Goldstone mode.

Returning to the case at hand, apart from the two constraints, this is an inhomogeneous set of ordinary differential equations which are sourced by quadratic terms in $\delta\rho_{\ast(0)}$. An important observation to make here is that the homogeneous part of these equations is also solved by the perturbations $\delta U_{\#(2)}$, $\delta g_{\#(2)}$, $\delta a_{\#(2)}$, $\delta \phi_{\#(2)}$ defined by normal phase expansions of \eqref{eq:normal_exp} defined by any choice of expansion parameter $\varepsilon_\#$. For such a solution of the homogeneous part we will have a trivial contribution to $\delta\tilde{\rho}_{(2)}$.

The final goal is the construction of a linear superposition of perturbations coming from the first corrections of the broken phase branch \eqref{eq:broken_exp} and normal branch \eqref{eq:normal_exp} such that it doesn't carry any net charge or entropy density. More precisely, after setting the value of the perturbative parametres equal $\varepsilon_\ast=\varepsilon_\#$, we have,
\begin{align}\label{eq:construct_sol}
\delta\tilde{U}_{(2)}&=\delta U_{\ast(2)}-\delta U_{\#(2)},\quad \delta\tilde{g}_{(2)}=\delta g_{\ast(2)}-\delta g_{\#(2)}\,\nn
\delta\tilde{a}_{(2)}&=\delta a_{\ast(2)}-\delta a_{\#(2)}-\delta\mu_{\ast(2)}+\delta\mu_{\#(2)}\,,\nn
\delta\tilde{\phi}_{(2)}&=\delta \phi_{\ast(2)}-\delta \phi_{\#(2)}\,,
\end{align}
with the parameters $\delta T_{\#(2)}$ and $\delta \mu_{\#(2)}$ such that the total perturbation does not carry any net charge or entropy density. Moreover, from the form of the solution \eqref{eq:construct_sol}, we can see that the source for scalar field $\phi$ in the perturbation is,
\begin{align}\label{eq:scalar_def}
\delta\tilde{\phi}_{(s)(2)}=\delta\phi_{(s)\ast (2)}-\delta\phi_{(s)\# (2)}\,.
\end{align}
Therefore, when we look for quasi-normal modes when trying to compute the gap, we will need to set $\delta\tilde{\phi}_{(s)(2)}=0$ in the perturbation.

Notice that in order to obtain a source free perturbation for the gauge field, we also performed a gauge transformation with parameter $\Lambda=(-\delta\mu_{\ast(2)}+\delta\mu_{\#(2)})\,\left(t+S(r)\right)$ with $S(r)$ as in \eqref{eq:qnm_ansatz}. This step will not be possible in section \ref{sec:superfluids} where we will be dealing with superfluids and the bulk 1-form field will become massive due to the spontaneous symmetry breaking. As we will see, the non-trivial asymptotics of the bulk 1-form field will be rather related to the field theory Goldstone which we expect to be involved in the case of a superfluid at finite chemical potential, with no particle-hole symmetry.

In general we can write variations of the entropy, the charge densities and the scalar VEV in terms of varations of the temperature, the chemical potential and the scalar source according to,
\begin{align}\label{eq:Tmu_relations}
\begin{pmatrix}
\delta s_\ast\\ \delta \varrho_\ast \\ \delta\langle \mathcal{O}_\phi\rangle_\ast
\end{pmatrix}
&=\frac{\varepsilon_\ast^2}{2}
\begin{pmatrix}
T_c^{-1}\,c_\mu^\ast & \xi^\ast & \nu_T^\ast\\
  \xi^\ast & \chi^\ast & \nu_\mu^\ast\\
 \nu_T^\ast & \nu_\mu^\ast & \nu_\phi^\ast
\end{pmatrix}
\begin{pmatrix}
\delta T_{\ast(2)} \\ \delta \mu_{\ast(2)} \\ \delta\phi_{(s)\ast (2)}
\end{pmatrix}
=\frac{\varepsilon_\ast^2}{2}\,\,
\begin{pmatrix}
\mathbf{\Xi}_\ast & \mathbf{\nu}_\ast \\
\mathbf{\nu}_{\ast}^{T} & \nu^\ast_\phi
\end{pmatrix}
\begin{pmatrix}
\delta T_{\ast(2)} \\ \delta \mu_{\ast(2)} \\ \delta\phi_{(s)\ast (2)}
\end{pmatrix}\,,
\nn
\begin{pmatrix}
\delta s_\#\\ \delta \varrho_\# \\ \delta\langle \mathcal{O}_\phi\rangle_\#
\end{pmatrix}
&=\frac{\varepsilon_\ast^2}{2}
\begin{pmatrix}
T_c^{-1}\,c_\mu^\# & \xi^\# & \nu_T^\#\\
  \xi^\# & \chi^\# & \nu_\mu^\#\\
 \nu_T^\# & \nu_\mu^\# & \nu_\phi^\#
\end{pmatrix}
\begin{pmatrix}
\delta T_{\#(2)} \\ \delta \mu_{\#(2)} \\ \delta\phi_{(s)\# (2)}
\end{pmatrix}
=\frac{\varepsilon_\ast^2}{2}\,\,
\begin{pmatrix}
\mathbf{\Xi}_\# & \mathbf{\nu}_\# \\
\mathbf{\nu}_{\#}^{T} & \nu^\#_\phi
\end{pmatrix}
\begin{pmatrix}
\delta T_{\#(2)} \\ \delta \mu_{\#(2)} \\ \delta\phi_{(s)\# (2)}
\end{pmatrix}\,,
\end{align}
where $\mathbf{\Xi}^\ast$ and $\mathbf{\Xi}^\#$ are the thermodynamic susceptibility matrices along the broken and normal phases correspondingly evaluated at the critical point. Using the above we can express temperature and chemical potential variations in terms of scalar deformations and entropy and charge densities according to,
\begin{align}\label{eq:Tmu_relations2}
\begin{pmatrix}
\delta T_{\ast(2)} \\ \delta \mu_{\ast(2)}
\end{pmatrix}&= \mathbf{\Xi}_\ast^{-1}\, \left(\delta\mathbf{\Phi}_{\ast (2)}
-\mathbf{\nu}_\ast\,  \delta\phi_{\ast (s) (2)}\right)\,,
\nn
\begin{pmatrix}
\delta T_{\#(2)} \\ \delta \mu_{\#(2)}
\end{pmatrix}&= \mathbf{\Xi}_\#^{-1}\, \left(\delta\mathbf{\Phi}_{\# (2)}
-\mathbf{\nu}_\#\,  \delta\phi_{\# (s) (2)}\right)\,.
\end{align}
In the above we have defined the vectors,
\begin{align}
\delta\mathbf{\Phi}_{\ast (2)}=
\begin{pmatrix}
\delta s_{\ast (2)} \\ \delta \varrho_{\ast (2)}
\end{pmatrix}
,\qquad
\delta\mathbf{\Phi}_{\# (2)}=
\begin{pmatrix}
\delta s_{\# (2)} \\ \delta \varrho_{\# (2)}
\end{pmatrix}\,,
\end{align}
which according to our earlier discussion, have to be equal to each other when constructing the time dependent perturbation. Finally, using equations \eqref{eq:Tmu_relations} and \eqref{eq:Tmu_relations2}, the variations of the scalar VEVs can be written as,
\begin{align}\label{eq:phi_vev_vars}
\delta\langle \mathcal{O}_\phi\rangle_{\ast(2)}&=\mathbf{\nu}^{\ast\,T} \mathbf{\Xi}_\ast^{-1}\, \delta\mathbf{\Phi}_{\ast (2)}+\left( \nu^\ast_\phi-\mathbf{\nu}_{\ast}^{T}\,\mathbf{\Xi}_\ast^{-1}\, \mathbf{\nu}_\ast\right)\, \delta \phi_{\ast (s) (2)}\,,\nn
\delta\langle \mathcal{O}_\phi\rangle_{\#(2)}&=\mathbf{\nu}^{\#\,T} \mathbf{\Xi}_\#^{-1}\, \delta\mathbf{\Phi}_{\# (2)}+\left(\nu_\phi^\#- \mathbf{\nu}_{\#}^{T}\,\mathbf{\Xi}_\#^{-1}\, \mathbf{\nu}_\# \right)\, \delta \phi_{\# (s) (2)}\,.
\end{align}
The above allows us to relate the thermodynamic susceptibilities between the two different ensembles through,
\begin{align}\label{eq:susc_relations}
\begin{pmatrix}
\left.\partial_s\langle \mathcal{O}_\phi\rangle\right|_{\varrho,\phi_{(s)}} \\ \left.\partial_\varrho\langle \mathcal{O}_\phi\rangle\right|_{s,\phi_{(s)}}
\end{pmatrix}
=\mathbf{\nu}^T\,\mathbf{\Xi}^{-1},\qquad \left.\partial_{\phi_{(s)}}\langle \mathcal{O}_\phi\rangle\right|_{s,\varrho}=\nu_\phi- \mathbf{\nu}^{T}\,\mathbf{\Xi}^{-1}\, \mathbf{\nu}\,.
\end{align}

From the horizon point of view, the zero net charge and entropy density translates to the boundary conditions\footnote{The reader might wonder about the fact that there is terms which involve different numbers of $\delta$ factors. We note that $\delta$ itself is of order zero in $\varepsilon_\ast$ and that depending on the origin of each factor, it can come with different powers of $\varepsilon_\ast$.},
\begin{align}
\delta \tilde{g}_{(2)}^{(0)}=0,\quad e^{2g_c^{(0)}}\left(Z_{c}^{(0)}\delta \tilde{a}_{(2)}^{\prime\,(0)}+\,a_c^{\prime\,(0)}\left( 2\,Z_{c}^{(0)}\,\delta \tilde{g}_{(2)}^{(0)}+\partial_\phi Z_c^{(0)}\,\delta\tilde{\phi}_{(2)}^{(0)}+\partial_{\rho^2} Z_c^{(0)}\,\delta\rho_{\ast(0)}^{(0)\,2}\right)\right)=0
\end{align}
 This will be useful in the next section where we employ the symplectic current to fix the term $\omega_{[2]}$ in the expansion \eqref{eq:timed_pert} and also show that $\omega_{[1]}=0$.
 
At this point, we understand the gapped mode in the bulk up to order $\varepsilon_\ast$ in the perturbative expansion. The ingredients we need are the static solutions of the broken and the normal phase bulk geometries. The final piece of information is the characteristic frequency $\omega_{[2]}$ which will be the task of the next subsection.

\subsection{The symplectic current}\label{sec:sympl_current}

In this section, we will employ the Crnkovic-Witten sympectic current to exract the final piece of information we are after, the frequency of the gapped mode close to the phase transition. To define it, we consider a generic classical Lagrangian field theory of a collection $\varphi^I$ of fields and two perturbative solutions $\delta_1\phi^I$ and $\delta_2\phi^I$ around a background $\phi^I_b$. If the Lagrangian density $\mathcal{L}(\varphi^I,\partial_\mu\varphi^I)$ can be written in terms of the fields and their first derivatives then the vector density,
\begin{align}\label{eq:scurrent_def}
P^\mu=\delta_1\phi^I\,\delta_2\left(\frac{\partial\mathcal{L}}{\partial \partial_\mu\phi^I} \right)-\delta_2\phi^I\,\delta_1\left(\frac{\partial\mathcal{L}}{\partial \partial_\mu\phi^I} \right)\,,
\end{align}
is divergence free,
\begin{align}\label{eq:div_free}
\partial_\mu P^\mu=0\,.
\end{align}
The symplectic current \eqref{eq:scurrent_def} is antisymmetric in field space and as such, when the second perturbative solution is infinitesimally close to the first one,  the symplectic form is expandable around zero.

As we will see, this observation is going to be particularly useful for us. In particular, the role of the background will be played by the black hole solution in the broken phase.  We will take the first of the two perturbations in \eqref{eq:scurrent_def} to be the static perturbation which is simply a derivative of the expansion \eqref{eq:broken_exp} with respect to $\varepsilon_\ast$ giving,
\begin{align}\label{eq:static_pert}
\delta g^\ast_{tt}&=-\varepsilon_\ast\,\delta U_{\ast(2)}-\frac{\varepsilon_\ast^3}{3!}\,\delta U_{\ast(4)}+\cdots,\nn
\delta g^\ast_{rr}&=-\varepsilon_\ast\frac{\delta U_{\ast(2)}}{U_c^2}-\frac{\varepsilon_\ast^3}{3!}\left(\frac{\delta U_{\ast(4)}}{U_c^2} -\frac{6\,\delta U_{\ast(2)}^2}{U_c^3}\right)+\cdots\,,\nn
\delta g^\ast_{ii}&=2\,\,e^{2g_c}\,\left(\varepsilon_\ast\,\delta g_{\ast(2)}+\frac{\varepsilon_\ast^3}{3!}\left(6\, \delta g_{\ast(2)}^2+\delta g_{\ast(4)}\right)+\cdots\right),\,\nn
\delta a^\ast_t&=\varepsilon_\ast\,\delta a_{\ast(2)}+\frac{\varepsilon_\ast^3}{3!}\,\delta a_{\ast(4)}+\cdots,\quad \delta\rho^\ast=\delta\rho_{\ast(0)}+\frac{\varepsilon_\ast^2}{2}\,\delta\rho_{\ast(2)}+\cdots\,,\nn
\delta\phi^\ast&=\varepsilon_\ast\,\delta\phi_{\ast(2)}+\frac{\varepsilon_\ast^3}{3!}\,\delta\phi_{\ast(4)}+\cdots\,.
\end{align}
The second perturbative solution will be the time dependent perturbation \eqref{eq:timed_pert} and the expansion parameter will naturally be given by $\varepsilon_\ast$. 

In order to obtain the symplectic current for the theory of equation \eqref{eq:bulk_action}, we first need to write it in a form where all the fields will appear with their first derivative at most. This will yield an equivalent action $\tilde{S}_{b}$ such that,
\begin{align}
\tilde{S}_{b}=S_{b}+S_{GH}=\int d^d x\,\mathcal{L}(\varphi^I,\partial_\mu\varphi^I)\,,
\end{align}
where we have introduced the Gibbons-Hawking term,
\begin{align}
S_{GH}=2\, \int_{\partial M}d^3 x\,\sqrt{-h}\,K\,,
\end{align}
and $\varphi^I$ include the metric along with the rest of the matter fields of our bulk theory. As usual, $K=\nabla_\mu n^\mu$ is the trace of the extrinsic curvature of the conformal boundary $\partial M$ and $n=dr/\sqrt{\mathcal{N}}$ is its normal one-form of unit norm. The integration measure is with respect to the induced metric $h_{\mu\nu}=g_{\mu\nu}-n_{\mu}n_{\nu}$.

For our bulk gravitational theory originating from \eqref{eq:bulk_action}, the sympectic current will read,
\begin{align}\label{eq:scurrent_bulk}
P^\mu=&\delta_1 g_{\alpha\beta}\,\delta_2\left(\frac{\partial\mathcal{L}}{\partial \partial_\mu g_{\alpha\beta}} \right)-\delta_2 g_{\alpha\beta}\,\delta_1\left(\frac{\partial\mathcal{L}}{\partial \partial_\mu g_{\alpha\beta}} \right)+\delta_1 A_\alpha\,\delta_2\left(\frac{\partial\mathcal{L}}{\partial \partial_\mu A_\alpha} \right)-\delta_2 A_\alpha\,\delta_1\left(\frac{\partial\mathcal{L}}{\partial \partial_\mu A_\alpha} \right)\notag\\
&+\delta_1 \phi\,\delta_2\left(\frac{\partial\mathcal{L}}{\partial \partial_\mu \phi} \right)-\delta_2 \phi\,\delta_1\left(\frac{\partial\mathcal{L}}{\partial \partial_\mu \phi} \right)+\delta_1 \rho\,\delta_2\left(\frac{\partial\mathcal{L}}{\partial \partial_\mu \rho} \right)-\delta_2 \rho\,\delta_1\left(\frac{\partial\mathcal{L}}{\partial \partial_\mu \rho} \right)\,.
\end{align}
In Appendix \ref{app:sympl_current_terms} we evaluate the derivatives of the bulk action density with respect to the partial derivatives of our fields.

The next step is to evaluate the symplectic current \eqref{eq:scurrent_bulk} for the pair of perturbations \eqref{eq:static_pert} and \eqref{eq:timed_pert} around the expanded broken phase background \eqref{eq:broken_exp}.  Before we do that, it is good to understand how we can benefit from considering the condition \eqref{eq:div_free} in our setup. Since we are Fourier expanding our time dependent modes, we will write the non-trivial components as,
\begin{align}
\mathcal{P}^t=e^{-i\omega (t+S(r))}\,P^{t}(r),\quad \mathcal{P}^r=e^{-i\omega (t+S(r))}\,P^r(r)\,.
\end{align}
The continuity equation \eqref{eq:div_free} then becomes,
\begin{align}\label{eq:scurrent_fourier}
-i\omega\,(P^t+S^\prime\,P^r)+P^{r\prime}=0\,.
\end{align}
Considering now the expansion in $\varepsilon_\ast$, we can write,
\begin{align}
P^t&=\varepsilon_\ast\,P^t_{(1)}+\varepsilon_\ast^2\,P^t_{(2)}+\cdots\,,\nn
P^r&=\varepsilon_\ast\,P^r_{(1)}+\varepsilon_\ast^2\,P^r_{(2)}+\cdots\,,
\end{align}
from where we see that at leading order \eqref{eq:scurrent_fourier} implies,
\begin{align}\label{eq:scurrent_fourier_1}
P^{r\prime}_{(1)}=0\,.
\end{align}
We therefore see that the term involving the time derivative in \eqref{eq:div_free}, is subleading in the $\varepsilon_\ast$ expansion. Keeping up to order $\varepsilon_\ast$ in the radial component of the symplectic current we obtain,
\begin{align}
P^{r}_{(1)}=-i\omega_{[1]}\,e^{2g_c}\, U_c\,S^\prime \,\delta\rho_{\ast(0)}^2\,.
\end{align}
After integrating \eqref{eq:scurrent_fourier_1} from the horizon up to infinity, we can show that $\omega_{[1]}=0$, as promised. This implies that after moving on to higher order in the perturbative expansion we will have that,
\begin{align}\label{eq:scurrent_fourier_2}
P^{r\prime}_{(2)}=0\,,
\end{align}
which will allow us to determine $\omega_{[2]}$, as we will see shortly. Expanding the sympectic current at next to leading order we obtain,
\begin{align}\label{eq:pr2}
P^{r}_{(2)}&=e^{2g_c}\,\left( -i\omega_{[2]}\,U_c\,S^\prime \,\delta\rho_{\ast(0)}^2+Z_c\,\left(\delta\tilde{a}_{(2)}\,\delta a_{\ast(2)}^\prime-\delta\tilde{a}^\prime_{(2)}\,\delta a_{\ast(2)}\right)\right)\nn
+&2 e^{2g_c}\, \left(\delta g_{\ast(2)}\,\left(Z_c\,\delta\tilde{a}_{(2)}\,a_c^\prime-\delta \tilde{U}_{(2)}^\prime\right)-\delta \tilde{g}_{(2)}\,\left(Z_c\,\delta a_{\ast(2)}\,a_c^\prime-\delta U_{\ast(2)}^\prime \right)\right)\nn
+&2\, U_c\,e^{2g_c}\,\left(\delta\tilde{g}_{(2)}\left( 2\, \delta g_{\ast(2)}^\prime +\delta\rho_{\ast(0)}\,\delta\rho_{\ast(0)}^\prime\right)-\delta g_{\ast(2)}\left( 2\, \delta\tilde{g}_{(2)}^\prime +\delta\rho_{\ast(0)}\,\delta\rho_{\ast(0)}^\prime\right)\right)\nn
-&\frac{1}{2}e^{2g_c}\,U_c\,\left(\delta\rho_{\ast(0)}^\prime\,\left(\delta\tilde{\rho}_{(2)}-\delta\rho_{\ast(2)} \right)- \delta\rho_{\ast(0)}\,\left(\delta\tilde{\rho}^\prime_{(2)}-\delta\rho^\prime_{\ast(2)} \right)\right)\nn
+&e^{2g_c}\left(-\delta U_{\ast (2)}\,\delta\tilde{\phi}_{(2)}+\delta \tilde{U}_{(2)}\,\delta\phi_{\ast (2)}+2\,U_c\,\left( -\delta g_{\ast (2)}\,\delta\tilde{\phi}_{(2)}+\delta\tilde{g}_{(2)}\,\delta\phi_{\ast (2)}\right) \right)\,\phi^\prime_c\nn
+&e^{2g_c}\,U_c\,\left(\delta\phi_{\ast (2)}\,\delta\tilde{\phi}_{(2)}^\prime-\delta\tilde{\phi}_{(2)}\,\delta\phi_{\ast (2)}^\prime \right)\nn
-&e^{2g_c}\delta U_{\ast(2)}\,\left( 2\, \delta \tilde{g}_{(2)}^\prime +\delta\rho_{\ast(0)}\,\delta\rho_{\ast(0)}^\prime\right)+e^{2g_c}\,\delta \tilde{U}_{(2)}\,\left( 2\, \delta g_{\ast(2)}^\prime +\delta\rho_{\ast(0)}\,\delta\rho_{\ast(0)}^\prime\right)\nn
-&\partial_\phi Z_c\,e^{2g_c}\,\left( \delta a_{\ast(2)}\,\delta\tilde{\phi}_{(2)}-\delta\tilde{a}_{(2)}\,\delta\phi_{\ast(2)}\right)\,a_c^\prime-\partial_{\rho^2} Z_c\,e^{2g_c}\,\left(\delta a_{\ast(2)}-\delta\tilde{a}_{(2)} \right)\,\delta\rho_{\ast(0)}^2\,a_c^\prime\,.
\end{align}
Integrating equation \eqref{eq:scurrent_fourier_2} from the horizon to infinity yields a term on the horizon and a term evaluated at the conformal boundary which have to be equal to each other. The expression we obtain reads,
\begin{align}
&-i\,\omega_{[2]}\,e^{2g_c^{(0)}}\,\delta\rho_{\ast(0)}^{(0)\,2}-2\,e^{2g_c^{(0)}}\,\delta g_{\ast(2)}\,\delta \tilde{U}_{(2)}^{\prime\,(0)}\nn
&+\delta\tilde{a}_{(2)}^{(0)}\,e^{2g_c^{(0)}}\left(Z_{c}^{(0)}\delta a_{\ast(2)}^{\prime\,(0)}+\,a_c^{\prime\,(0)}\left( 2\,Z_{c}^{(0)}\,\delta g_{\ast(2)}^{(0)}+\partial_\phi Z_c^{(0)}\,\delta\phi_{\ast(2)}^{(0)}+\partial_{\rho^2} Z_c^{(0)}\,\delta\rho_{\ast(0)}^{(0)\,2}\right)\right)=\nn
&-\left(2\Delta_\phi-3 \right)\delta\phi_{(s)\ast (2)}\,\left( \delta\phi_{(v)\ast(2)} - \delta\phi_{(v)\#(2)}\right)\,,
\end{align}
where we imposed the absence of scalar sources on the boundary by setting $\delta\phi_{(s)\ast(2)}=\delta\phi_{(s)\#(2)}$ in equation \eqref{eq:scalar_def}. After recognising the variations of the horizon charge and entropy densities \eqref{eq:cs_horizon}, we can write,
\begin{align}
&\qquad -i\,\omega_{[2]}\,\frac{s_{c}}{4\pi}\,\delta\rho_{\ast(0)}^{(0)\,2}=\nn
& \delta s_{\ast(2)}\,(\delta T_{\ast(2)}-\delta T_{\#(2)})+\delta\varrho_{\ast(2)} \,(\delta \mu_{\ast(2)}-\delta \mu_{\#(2)})-\delta\phi_{(s)\ast(2)}\,\left(\delta\langle\mathcal{O}_\phi\rangle_{\ast(2)}-\delta\langle\mathcal{O}_\phi\rangle_{\#(2)} \right)\,,
\end{align}
with the variations of the scalar VEVs defined according to \eqref{eq:vev_vars}. After introducing appropriate factors of $\varepsilon_\ast$,  setting, 
\begin{align}\label{eq:Phi_def}
\delta\mathbf{\Phi}=\frac{\varepsilon_\ast^2}{2}\,\delta\mathbf{\Phi}_{\ast (2)}\,,\qquad \delta\phi_{(s)}=\frac{\varepsilon_\ast^2}{2}\,\delta\phi_{(s)\ast(2)}\,.
\end{align}
and using equations \eqref{eq:Tmu_relations2} and \eqref{eq:phi_vev_vars}, the frequency of our Higgs takes the remarkably simple form,
\begin{align}\label{eq:Higgs_gap_neutral}
\omega=i\frac{8\,\Delta E}{\varpi}\,.
\end{align}
In the above, we have defined the horizon quantity,
\begin{align}
\varpi=\frac{s_c\,\rho^{(0)\,2}}{4\,\pi}\,,
\end{align}
as well as 
\begin{align}\label{eq:energy_gap}
\Delta E=&\frac{1}{2}\delta\mathbf{\Phi}^T\,\left(\mathbf{\Xi}_{\ast}^{-1}-\mathbf{\Xi}_{\#}^{-1} \right)\delta\mathbf{\Phi}-\delta\mathbf{\Phi}^T\left(\mathbf{\Xi}_{\ast}^{-1}\mathbf{\nu}_\ast-\mathbf{\Xi}_{\#}^{-1}\mathbf{\nu}_\# \right)\delta\phi_{(s)}\nn
&-\frac{1}{2}\left(\nu^\ast_\phi-\mathbf{\nu}_\ast^T\mathbf{\Xi}_\ast^{-1}\mathbf{\nu}_\ast-\nu^\#_\phi+\mathbf{\nu}_\#^T\mathbf{\Xi}_\#^{-1}\mathbf{\nu}_\# \right)\,\delta\phi_{(s)}^2\,.
\end{align}
As we show in Appendix \ref{app:thermo}, $\Delta E$ is the energy density difference between the normal and broken phase thermal states at entropy density $s+\delta s_\ast$, charge density $\varrho+\delta\varrho_\ast$ and scalar deformation parameter $\phi_{(s)}+\delta\phi_{(s)\ast}$.  As we would expect, the frequency of the Higgs mode \eqref{eq:energy_gap} is in the lower half of the complex plane as long as the energy of the broken phase black holes is lower than then normal phase in the microcanonical ensemble.

As we will see later, apart from the energy, there will be other quantities which will have to compare between the two phases. For this reason, if a thermodynamic quantity $\mathcal{O}$ can be expressed in terms of the entropy density, the charge density and the scalar deformation, $\Delta\mathcal{O}$ will denote the difference,
\begin{align}\label{eq:Delta_def}
\Delta\mathcal{O}=\mathcal{O}_\ast(s_c+\delta s_\ast,\varrho+\delta\varrho_\ast,\phi_{(s)}+\delta\phi_{(s)\ast})-\mathcal{O}_\#(s_c+\delta s_\ast,\varrho+\delta\varrho_\ast,\phi_{(s)}+\delta\phi_{(s)\ast})\,,
\end{align}
with $\mathcal{O}_\ast$ and $\mathcal{O}_\#$ the values of $\mathcal{O}$ in the broken and normal phase respectively. Moreover, $s_c$ is the entropy at the critical point and is set by $\rho$ and $\phi_{(s)}$. In our work we will only need the leading order approximation in $\varepsilon_\ast$ where as we know $\delta s_\ast\sim \delta\varrho_\ast\sim \delta\phi_{(s)\ast}\sim O(\varepsilon_\ast^2)$.

\section{Charged Superfluids}\label{sec:superfluids}
In this section, we will examine the case in which the broken phase describes a superfluid.  To achieve this, the bulk theory will need to contain a complex scalar field $\psi$ with a $U(1)$ gauged symmetry which will drive the phase transition. For concreteness, we will consider the bulk action,
\begin{align}\label{eq:bulk_action_charged}
S_{b}=\int d^4x\,\sqrt{-g}\,\left( R-V-\frac{1}{2}\left(\partial\phi \right)^2-\frac{1}{2}\left(D \psi \right)^2-\frac{Z}{4}F^2\right)\,,
\end{align}
with $D_\mu \psi=\left(\nabla_\mu+i\,q\,A_\mu\right)\,\psi$. 

In the superfluid phase, the operator $\mathcal{O}_\psi$ which is dual to the bulk field $\psi$ takes a non-trivial VEV yielding a corresponding non-trivial bulk field $\psi$. In order to proceed, we will perform a field transformation to write the complex scalar field in a polar decomposition according to,
\begin{align}\label{eq:polar_decomp}
\psi=\rho\,e^{i\,q\,\theta}\,,
\end{align}
with $\rho>0$ and $0\leq \theta <2\pi/q$. The above transformation brings the bulk action to the form,
\begin{align}\label{eq:bulk_action_charged_V2}
S_b=\int d^4x\,\sqrt{-g}\,\left( R-V-\frac{1}{2}\left(\partial\phi \right)^2-\frac{1}{2}\left(\partial \rho \right)^2-\frac{Z}{4}F^2-\frac{q^2}{2}\rho^2\,B^2\right)\,,
\end{align}
where we have set $B=A+\partial\theta$ and therefore $F=dB$. This form is very similar to the bulk action \eqref{eq:bulk_action} that we considered in the previous sections for  neutral scalar fields which are dual to real order parameters. The crucial difference is that in the broken phase, the 1-form field does not enjoy gauge invariance due to the Stueckelberg mechanism.

We would now like to consider the asymptotic expansion of 1-form field $B_\alpha$ fluctuations. For the components along the conformal boundary directions we obtain,
\begin{align}\label{eq:uv_bexp}
\delta B_\alpha=\frac{\partial_\alpha \delta\theta_{(s)}}{(r+R)^{3-2\,\Delta_\rho}}+\cdots+\delta A_\alpha+\partial_\alpha\delta\theta_{(v)}+\cdots+\frac{\delta j_\alpha}{r+R}+\cdots\,,
\end{align}
with the radial component $B_r$ completely determined by this. The asymptotic expansion of the 1-form field $B$ is therefore fixed by a scalar function $\theta_{(s)}$ and the two 1-forms  $A_\alpha+\partial_\alpha\,\theta_{(v)}$ and $j_\alpha$. 

The function $\theta_{(v)}$ and the parameter $\rho_{(v)}$ that appears in the expansion \eqref{eq:uv_bcs} of the modulus $\rho$ of the complex field $\psi$ combine together to  parametrise the VEV,
\begin{align}
\langle\mathcal{O}_\psi\rangle=\frac{1}{2}(2\Delta_\rho -3)\,\rho_{(v)}e^{i\,q\,\theta_{(v)}}\,,
\end{align}
from which we obtain the perturbative epxression
\begin{align}\label{eq:o_psi_var}
\delta\langle\mathcal{O}_\psi\rangle=\langle\mathcal{O}_\psi\rangle_b\,i\,q\,\delta\theta_{(v)}+e^{i\,\arg(\langle\mathcal{O}_\psi\rangle_b)}\, \delta|\langle\mathcal{O}_\psi\rangle|\,,
\end{align}
 in terms of the VEV of the thermal state $\langle\mathcal{O}_\psi\rangle_b$. The combination $\delta A_\alpha+\delta\partial_\alpha\theta_{(v)}$ is then a gauge invariant combination of the perturbative external source $\delta A_\alpha$ and the phase of the VEV $\delta\theta_{(v)}$. In the absence of external sources, the constants of integration $j_\alpha$ are directly related to VEV of the boundary charge density $\langle J\rangle_\alpha$ satisfying the continuity equation $\partial_\alpha j^\alpha=0$.

The treatment of the gapped mode is identical to that of the neutral scalar field with the gauge field $A$ replaced by the massive 1-form $B$. The whole argument would go through apart from one technical point of particular physical significance. This is related to the gauge transformation that we dicuss below equation \eqref{eq:construct_sol} and which removes the sources from the asymptotic of the gauge field $A$.  Since we cannot perform gauge transformations to $B$, the same treatment is not possible in this case. However, as we can see from the expansion \eqref{eq:uv_bexp},  a non zero value for $B$ on the conformal boundary can be treated as the Goldstone mode $\delta\theta_{(v)}$ as long as it is an exact form i.e. it can be expressed as the derivative of the function $\delta\theta_{(v)}$.  This is certainly possible for the perturbations we are studying in this paper since only the time and the radial components of the 1-form $B$ will be involved. The constant on the boundary coming from the time component can then always be expressed as a partial derivative of a phase $\delta\theta_{(v)}$ with respect to time.

To proceed, we introduce the Fourier modes of equation \eqref{eq:qnm_ansatz} according to,
\begin{align}
\delta B_\mu=e^{-i\omega (t+S(r))}\,\delta b_\mu(r)\,.
\end{align}
Another important ingredient in order to construct the perturbation of order $\varepsilon_\ast$ for the 1-form field is the discussion below equation \eqref{eq:timed_pert}. We can indeed form zero total charge linear combinations of the static perturbations $\delta a_{\ast(2)}$ and $\delta a_{\#(2)}$ as we did in the case of the neutral order parameter in order to obtain solutions of the time component of the equations of motion.  The important difference now is that the radial component is no longer a constraint which can be solved everywhere in the bulk by imposing charge continuity on the boundary. 

This is a signal that, in addition to our previous considerations, the phase $\delta\theta$ in the bulk is  going to get involved in the perturbation at order $\varepsilon$. More concretly, we can expand the bulk phase in $\varepsilon_\ast$ according to,
\begin{align}
\delta \tilde{\theta}=\varepsilon_\ast^{-1}\left( \delta \tilde{\theta}_{(0)}+\varepsilon_\ast^2\,\delta \tilde{\theta}_{(2)}+\cdots\right)\,,
\end{align}
where $\delta \tilde{\theta}_{(0)}$ is constant everywhere in the bulk and $\delta \tilde{\theta}_{(2)}$ an analytic function of the radial coordinate. The above implies that at order $\varepsilon_\ast$ the bulk perturbation for the one form field takes the form,
\begin{align}
\delta \tilde{b}_{t(2)}&=\delta a_{\ast(2)}-\delta a_{\#(2)}-i\omega_{[2]}\,\delta \tilde{\theta}_{(0)}\,,\notag\\
\delta \tilde{b}_{r(2)}&=-i\,S^\prime\,\omega_{[2]}\,\delta \tilde{\theta}_{(0)}+\delta \tilde{\theta}_{(2)}^\prime\,.
\end{align}
A crucial point to note here is that the terms related to $\delta \tilde{\theta}_{(0)}$ and $\delta \tilde{\theta}_{(2)}$  enter the time component of the equation of motion of $\delta B$ at order $\varepsilon_\ast^3$. This shows that the above solves the time dependent equation at order $\varepsilon_\ast$ even after the addition of the terms which are an exact form.

On the other hand, as one can easily see, both $\delta \tilde{\theta}_{(0)}$ and $\delta \tilde{\theta}_{(2)}$ will enter the radial component of the equation of motion of $\delta B$ at order $\varepsilon_\ast^3$. Expanding this equation close to the horizon with the boundary conditions given in \eqref{eq:nh_bcs}, \eqref{eq:gen_exp} and \eqref{eq:nh_reg} we obtain,
\begin{align}\label{eq:bulk_phase_sol}
\delta\tilde{\theta}_{(0)}&=\frac{1}{q^2\,\delta\rho_{\ast(0)}^{(0)\,2}}\,\left(Z_{c}^{(0)}\delta \tilde{a}_{(2)}^{\prime\,(0)}+\,a_c^{\prime\,(0)}\left( 2\,Z_{c}^{(0)}\,\delta \tilde{g}_{(2)}^{(0)}+\partial_\phi Z_c^{(0)}\,\delta\tilde{\phi}_{(2)}^{(0)}+\partial_{\rho^2} Z_c^{(0)}\,\delta\tilde{\rho}_{(0)}^{(0)\,2}\right)\right)\\\notag
&=\frac{4\pi}{s_c\,q^2\,\delta\rho_{\ast(0)}^{(0)\,2}}\,\left(\delta\varrho^{(0)}_{\ast(2)}-\delta\varrho_{\#(2)}\right)=\frac{4\pi}{s_c\,q^2\,\delta\rho_{\ast(0)}^{(0)\,2}}\,\left(\delta\varrho^{(0)}_{\ast(2)}-\delta\varrho_{\ast(2)}\right)\,.
\end{align}

To obtain the first equality in the second line of \eqref{eq:bulk_phase_sol},  we used equation \eqref{eq:cs_horizon} which relates the charge density perturbation of the normal phase $\delta\varrho_{\#(2)}$ to the flux density of the gauge field at the horizon. At the same time, we have defined $\delta\varrho^{(0)}_{\ast(2)}$, the perturbation of a horizon charge density related to the 1-form field $B$ in the broken phase. In the case of superfluids, an important point is that this is not equal to the charge density of the dual field theory which needs to be read off from the bulk fields after expanding close to the conformal boundary. Finally, to obtain the last equality, we have used that the two perturbations that we obtain from the static backgrounds have to be chosen so that $\delta\varrho_{\#(2)}=\delta\varrho_{\ast(2)}$. We see that the bulk constant $\delta\tilde{\theta}_{(0)}$ is directly proportional to the difference between the horizon and the conformal boundary charge densities.

We now turn our attention to the symplectic current built from the background solution and the time dependent perturbation of our gapped mode. We follow the same logic as in section \ref{sec:sympl_current} for the neutral case. In fact, the bulk actions \eqref{eq:bulk_action} and \eqref{eq:bulk_action_charged_V2} have identical kinetic terms and they will therefore yield the same form for the symplectic current. Keeping terms up to order $\varepsilon^2_\ast$ yields the same form \eqref{eq:pr2} for the radial component.

Following the argument of section \ref{sec:sympl_current}, we equate the value of the radial component $P_{(2)}^r$ on the horizon to its value at the conformal boundary. In this case however, there is a non-zero contribution from the conformal boundary due to the non-trivial asymptotics of the vector field component $\delta \tilde{b}_{t(2)}$.  This allows us to write,
\begin{align}
&-i\,\omega_{[2]}\,e^{2g_c^{(0)}}\,\delta\rho_{\ast(0)}^{(0)\,2}-i\omega_{[2]}\,\delta\tilde{\theta}^{(0)}\,\left(\delta\varrho^{(0)}_{\ast(2)}-\delta\varrho_{\ast(2)}\right)\nn
&= \delta s_{\ast(2)}\,(\delta T_{\ast(2)}-\delta T_{\#(2)})+\delta\varrho_{\ast(2)} \,(\delta \mu_{\ast(2)}-\delta \mu_{\#(2)})-\delta\phi_{(s)\ast(2)}\,\left(\delta\langle\mathcal{O}_\phi\rangle_{\ast(2)}-\delta\langle\mathcal{O}_\phi\rangle_{\#(2)} \right)\,.
\end{align}
After using equation \eqref{eq:bulk_phase_sol} for the zero order perturbation of the phase $\delta\tilde{\theta}_{(0)}$, we obtain the simple result,
\begin{align}\label{eq:gap_super}
\omega=i\,\frac{8\,\Delta E}{\varpi}\,,
\end{align}
where we have defined the transport coefficient,
\begin{align}
\varpi=\frac{s_c\,\rho^{(0)\,2}}{4\,\pi}+\frac{16\pi}{s_c\,q^2\,\rho^{(0)\,2}}\left(\varrho^{(0)}-\varrho\right)^2\,,
\end{align}
and $\Delta E$ remains the energy difference of equation \eqref{eq:energy_gap}. This transport coefficient is intimately connected to the dynamics of the Higgs mode. A certain way to understand it is through the decay rate of equation \eqref{eq:gap_super}. Finally, by using the equation of motion for the bulk one form field, it is easy to show that close to the phase transitions $\varrho^{(0)}-\varrho\sim\delta\rho^2\sim\varepsilon^2$ giving that $\varpi\sim\varepsilon^2$.

\section{Green's Functions of Scalar Operators}\label{sec:greens}
In this section we wish to study the effect of the Higgs mode we constructed in the previous section on the linear response of the system near the critical point. For this reason, we will need to introduce boundary sources for the scalar fields in the time dependent perturbation that we constructed in section \ref{sec:t_pert}.

As we will see shortly, the scalar sources will have to be turned on at orders $\delta\rho_{(s)}\sim O(\varepsilon^2_\ast)$ and $\delta\phi_{(s)}\sim O(\varepsilon_\ast)$ in the $\varepsilon_\ast$ expansion of equation \eqref{eq:timed_pert}. Including sources does not change the arguments around the construction of sections \ref{sec:t_pert} and \ref{sec:superfluids}. As we have seen there, at next to leading order in $\varepsilon_\ast$, the perturbation can is simply a linear combination of a variation of the broken and the unbroken phase black hole backgrounds. The former variation is fixed by how we move on the phase diagram and does not allow us to introduce new sources in the perturbation.

This shows that the only way to introduce perturbative source for the scalar $\phi$, which are not present in the background thermal state, is through the variation of the normal phase branch of the black holes $\delta\phi_{\# (2)}$ in equation \eqref{eq:construct_sol}. However, for the scalar which is relevant to the amplitude of the order parameter $\langle\mathcal{O}_\rho\rangle$, the source will need to be introduced in the $O(\varepsilon^3)$ term of equation \eqref{eq:timed_pert}. This can be achieved through the asymptotics of the bulk function $\delta\tilde{\rho}_{(2)}$.

We will start our treatment by considering a source $\delta\tilde{\rho}_{(s)}$ for the operator $\mathcal{O}_\rho$. As described above, this appears as a constant of integration in the asymptotic expansion,
\begin{align}
\delta\tilde{\rho}_{(2)}=\delta\tilde{\rho}_{(s)(2)}\,(r+R)^{\Delta_\rho -3}+\cdots+\delta\tilde{\rho}_{(v)(2)}\,(r+R)^{-\Delta_\rho}+\cdots\,.
\end{align}
for the function $\delta\tilde{\rho}_{(2)}$ defined through equation \eqref{eq:timed_pert}. This proves our earlier statement that the source for $\mathcal{O}_\rho$ should be of order $\varepsilon_\ast^2$ with $\delta\tilde{\rho}_{(s)}=\frac{\varepsilon_\ast^2}{2}\,\delta\tilde{\rho}_{(s)(2)}$. By equating the value of the radial component of the symplectic current \eqref{eq:pr2} on the horizon and the conformal boundary we obtain,
\begin{align}\label{eq:soure_rho}
&-i\,\omega_{[2]}\,e^{2g_c^{(0)}}\,\delta\rho_{\ast(0)}^{(0)\,2}-i\omega_{[2]}\,\delta\tilde{\theta}^{(0)}\,\left(\delta\varrho^{(0)}_{\ast(2)}-\delta\varrho_{\ast(2)}\right)\nn
&= \delta s_{\ast(2)}\,(\delta T_{\ast(2)}-\delta T_{\#(2)})+\delta\varrho_{\ast(2)} \,(\delta \mu_{\ast(2)}-\delta \mu_{\#(2)})-\delta\phi_{(s)\ast(2)}\,\left(\delta\langle\mathcal{O}_\phi\rangle_{\ast(2)}-\delta\langle\mathcal{O}_\phi\rangle_{\#(2)} \right)\nn
&\quad +\frac{1}{2}\,\delta\tilde{\rho}_{(s)(2)}\,\delta\langle\mathcal{O}_\rho\rangle_{\ast(0)}\,.
\end{align}
with $\delta\tilde{\theta}^{(0)}$ given once again by equation \eqref{eq:bulk_phase_sol}. The above allows us to relate the frequency $\omega=\varepsilon_\ast^2\,\omega_{[2]}$ to the source $\delta\tilde{\rho}_{(s)}$. On the other hand, the expansion \eqref{eq:vev_vars} for the VEVs of the operators $\mathcal{O}_\rho$ and $\mathcal{O}_\phi$ is enough to fix the retarded Green's function,
\begin{align}\label{eq:G_rho_rho}
G_{\mathcal{O}_\rho\mathcal{O}_\rho}(\omega)&=\frac{\delta\tilde{\langle\mathcal{O}_\rho\rangle}}{\delta\tilde{\rho}_{(s)}}=\frac{\delta \langle\mathcal{O}_\rho\rangle_\ast^2}{\varpi\,\left(- i\,\omega+\omega_{gap}\right)}\nn
=&\frac{(\Delta \langle\mathcal{O}_\rho\rangle )^2}{\varpi\,\left(- i\,\omega+\omega_{gap}\right)}\,,
\end{align}
where the symbol $\Delta$ is as defined in equation \eqref{eq:Delta_def}. Once again, our $\varepsilon_\ast$ of equation \eqref{eq:timed_pert} has singled out the pole which is relevant to the dominant Higgs mode. We expect a whole tower of quasi-normal modes with much larger decay rates and which will not influence the low energy dynamics of our system. In the above, we have defined the transport coefficient $\varpi$ and gap frequency $\omega_{gap}$ according to,
\begin{align}\label{eq:varpi_def}
\varpi=\frac{s_c\,\rho^{(0)\,2}}{4\,\pi}+\frac{16\pi}{s_c\,q^2\,\rho^{(0)\,2}}\left(\varrho^{(0)}-\varrho\right)^2,\qquad \omega_{gap}=-\frac{8\,\Delta E}{\varpi}\,.
\end{align}
This is the first Green's function we determine and as we anticipated it is dominated by a single simple pole at $\omega=-i\,\omega_{gap}$ agreeing with the gap of equation \eqref{eq:gap_super}. Notice, that equation \eqref{eq:G_rho_rho} allows us to write the expressions,
\begin{align}
\omega_{gap}&=\frac{(\Delta \langle\mathcal{O}_\rho\rangle )^2}{\chi_{\mathcal{O}_{\rho}\mathcal{O}_{\rho}}}\,\varpi^{-1}\,,\nn
\chi_{\mathcal{O}_{\rho}\mathcal{O}_{\rho}}&=-\frac{(\Delta \langle\mathcal{O}_\rho\rangle )^2}{8\,\Delta E}\,.
\end{align}
where $\chi_{\mathcal{O}_{\rho}\mathcal{O}_{\rho}}$ is the susceptibility of the operator $\mathcal{O}_{\rho}$.  The above allows us to write the Kubo formula,
\begin{align}
\frac{\varpi}{(\Delta \langle\mathcal{O}_\rho\rangle )^2}=\frac{1}{\chi_{\mathcal{O}_{\rho}\mathcal{O}_{\rho}}^2}\lim_{\omega\to 0}\frac{\mathrm{Im}\, G_{\mathcal{O}_\rho\mathcal{O}_\rho}(\omega)}{\omega}\,,
\end{align}
noting that the left hand side remains finite at the critical point, given the fact that $\varpi\sim \varepsilon^2$. This ratio is very closely related to the coefficient $\Gamma$ of \cite{RevModPhys.49.435} which the authors assume to remain finite at the critical point.

The next Green's function we can compute after having turned on the source for $\mathcal{O}_\rho$ is,
\begin{align}
G_{\mathcal{O}_\phi\mathcal{O}_\rho}(\omega)=\frac{\delta\tilde{\langle\mathcal{O}_\phi\rangle}}{\delta\tilde{\rho}_{(s)}}=2\,\frac{\delta \langle\mathcal{O}_\rho\rangle_\ast\,\left(\delta \langle\mathcal{O}_\phi\rangle_\ast-\delta \langle\mathcal{O}_\phi\rangle_\# \right)}{\varpi\,\left(- i\,\omega+\omega_{gap}\right)}\,,
\end{align}
where once again we used the relation \eqref{eq:soure_rho} for the source $\delta\tilde{\rho}_{(s)}$. In the current situation, we are not turning on the source $\delta\tilde{\phi}_{(s)}$ for the scalar operator. Therefore, in this case we simply have,
\begin{align}
\Delta \langle\mathcal{O}_\phi\rangle= \delta \langle\mathcal{O}_\phi\rangle_\ast-\delta \langle\mathcal{O}_\phi\rangle_\# \,,
\end{align}
allowing us to write,
\begin{align}\label{eq:g_phi_rho}
G_{\mathcal{O}_\phi\mathcal{O}_\rho}(\omega)=2\,\frac{\Delta \langle\mathcal{O}_\rho\rangle\,\Delta \langle\mathcal{O}_\phi\rangle}{\varpi\,\left(- i\,\omega+\omega_{gap}\right)}\,.
\end{align}
Once again, we see that the retarded Green's functions are determined by the transport coefficient $\varpi$ and thermodynamic properties of the broken and normal phases.

We now turn our attention to the source of the scalar operator $\mathcal{O}_\phi$. Introducing a source $\delta\tilde{\phi}_{(s)(2)}$ as in equation \eqref{eq:scalar_def} allows us to express the scalar source variation for the normal phase as,
\begin{align}\label{eq:Nphase_source_var}
\delta\phi_{(s)\#}=\delta\phi_{(s)\ast}-\delta\tilde{\phi}_{(s)}\,.
\end{align}

The final ingredient left in order to compute the retarded Green's function is a relation that we can obtain from the symplectic current \eqref{eq:pr2}. By evaluating it on the horizon and the conformal boundary we obtain the relation,
\begin{align}
&-i\,\omega_{[2]}\,e^{2g_c^{(0)}}\,\delta\rho_{\ast(0)}^{(0)\,2}-i\omega_{[2]}\,\delta\tilde{\theta}^{(0)}\,\left(\delta\varrho^{(0)}_{\ast(2)}-\delta\varrho_{\ast(2)}\right)\nn
&= \delta s_{\ast(2)}\,(\delta T_{\ast(2)}-\delta T_{\#(2)})+\delta\varrho_{\ast(2)} \,(\delta \mu_{\ast(2)}-\delta \mu_{\#(2)})-\delta\phi_{(s)\ast(2)}\,\left(\delta\langle\mathcal{O}_\phi\rangle_{\ast(2)}-\delta\langle\mathcal{O}_\phi\rangle_{\#(2)} \right)\nn
&\quad +\delta\tilde{\phi}_{(s)(2)}\,\delta\langle\mathcal{O}_\phi\rangle_{\ast (2)}\,.
\end{align}
After using equations \eqref{eq:Tmu_relations2} and \eqref{eq:phi_vev_vars} as well as the definitions \eqref{eq:Delta_def} and \eqref{eq:varpi_def}, the above takes the form,
\begin{align}\label{eq:phis_result}
2\,\varepsilon_\ast\,\delta\tilde{\phi}_{(s)}\,\Delta \langle \mathcal{O}_\phi\rangle&=-\varpi\,\left(i\,\omega-\omega_{gap}\right)\,.
\end{align}
The combination of this and relation \eqref{eq:vev_vars} allows us to compute the retarded Green's functions,
\begin{align}
G_{\mathcal{O}_\phi \mathcal{O}_\phi}(\omega)&=\frac{\delta \tilde{\langle \mathcal{O}_\phi \rangle}}{\delta\tilde{\phi}_{(s)}}=2\,\frac{\delta \langle \mathcal{O}_\phi \rangle_\ast-\delta \langle \mathcal{O}_\phi \rangle_\#}{\varepsilon_\ast\,\delta\tilde{\phi}_{(s)}}\nn
&=4\,\frac{\left(\Delta\langle \mathcal{O}_\phi\rangle\right)^2}{\varpi\,\left(-i\,\omega+\omega_{gap}\right)}+\left.\partial_{\phi_{(s)}}\langle\mathcal{O}_\phi\rangle_\# \right|_{s,\varrho}\,,
\end{align}
where the last term is computed in the normal phase, at the critical point. To see this relation, we have combined equations \eqref{eq:phi_vev_vars} and \eqref{eq:susc_relations} along with \eqref{eq:Nphase_source_var}. Finally, we can use the result of equation \eqref{eq:phis_result} to compute, 
\begin{align}\label{eq:g_rho_phi}
G_{\mathcal{O}_\rho \mathcal{O}_\phi}(\omega)&=\frac{\delta \tilde{\langle \mathcal{O}_\rho \rangle}}{\delta\tilde{\phi}_{(s)}}=\frac{\delta \langle \mathcal{O}_\rho \rangle_\ast}{\varepsilon_\ast\,\delta\tilde{\phi}_{(s)}}\notag\\
&=2\,\frac{\Delta\langle \mathcal{O}_\phi\rangle \,\Delta \langle \mathcal{O}_\rho \rangle}{\varpi\,\left(-i\,\omega+\omega_{gap}\right)}\,.
\end{align}
From the form of the Green's functions in equations \eqref{eq:g_phi_rho} and \eqref{eq:g_rho_phi}, it is trivial to check the validity of the Onsager relation $G_{\mathcal{O}_\rho \mathcal{O}_\phi}(\omega)=G^\ast_{\mathcal{O}_\phi \mathcal{O}_\rho}(-\omega)$.

It is interesting to examine the behaviour of the boundary theory Goldstone mode which is represented by the angle $\delta\theta_{(v)}$ of equation \eqref{eq:o_psi_var}. This can be read off from the asymptotics of the time component of the 1-form field $\delta\tilde{b}_t$ and interpreting the constant term as the time derivative $\partial_t\delta\theta_{(v)}$. This gives,
\begin{align}\label{eq:phase_shift}
\delta\tilde{\theta}_{(v)(0)}=\delta\tilde{\theta}_{(0)}+\frac{2\,i}{\omega}\,\left(\delta\mu_{\ast}-\delta\mu_{\#} \right)\,,
\end{align}
with the variation of the normal phase chemical potential $\delta\mu_{\# (2)} $ being fixed by equation \eqref{eq:Tmu_relations2}.

The VEV of the operator $\mathcal{O}_\rho$ is related to the amplitude of the order parameter, according to the decomposition of fluctuations \eqref{eq:o_psi_var}. More specifically, in the broken phase, where the polar coordinate decomposition \eqref{eq:polar_decomp} makes sense, we have,
\begin{align}
\delta \langle\mathcal{O}_\rho\rangle=2\,\delta|\langle\mathcal{O}_\psi\rangle|=\frac{1}{|\langle\mathcal{O}_\psi\rangle_b|}\,\left(\langle\mathcal{O}_{\bar{\psi}}\rangle_b\,\delta\langle\mathcal{O}_\psi\rangle+\langle\mathcal{O}_\psi\rangle_b\,\delta\langle\mathcal{O}_{\bar{\psi}}\rangle\right)\,.
\end{align}
It is interesting to also consider the operator $\mathcal{O}_Y$ which is related to the phase of the order parameter and therefore the Goldstone mode of the theory. For the fluctuations of this operator we can write,
\begin{align}\label{eq:phase_operator}
\delta\langle \mathcal{O}_Y\rangle=q\,\langle\mathcal{O}_\rho\rangle_b\,\delta\theta_{(v)}=\frac{1}{i\,|\langle\mathcal{O}_\psi\rangle_b|}\,\left(\langle\mathcal{O}_{\bar{\psi}}\rangle_b\,\delta\langle\mathcal{O}_\psi\rangle-\langle\mathcal{O}_\psi\rangle_b\,\delta\langle\mathcal{O}_{\bar{\psi}}\rangle\right)\,.
\end{align}

Given the expression \eqref{eq:phase_shift} for the phase, we can easily compute the retarded Green's functions measuring the response of the operator $\mathcal{O}_Y$ against sources for the operators $\langle\mathcal{O}_\rho\rangle$ and $\langle\mathcal{O}_\phi\rangle$,
\begin{align}
G_{\mathcal{O}_Y \mathcal{O}_\phi}(\omega)&=\frac{\delta\tilde{\langle \mathcal{O}_Y\rangle}}{\delta \tilde{\phi}_{(s)}}=\frac{2\,q\,\Delta\langle\mathcal{O}_\rho\rangle\,\Delta\langle\mathcal{O}_\phi\rangle}{\omega}\,\frac{\omega\,\vartheta+2\,i\,\Delta\mu}{\varpi\,\left(-i\,\omega+\omega_{gap}\right)}+\frac{2\,q\,\Delta\langle\mathcal{O}_\rho\rangle}{i\,\omega}\,\left.\frac{\partial\mu_\#}{\partial\phi_{(s)}}\right|_{s,\varrho}\,,\nn
G_{\mathcal{O}_Y \mathcal{O}_\rho}(\omega)&=\frac{\delta\tilde{\langle \mathcal{O}_Y\rangle}}{\delta \tilde{\rho}_{(s)}}=\frac{q\,\Delta\langle\mathcal{O}_\rho\rangle^2}{\omega}\,\frac{\omega\,\vartheta+2\,i\,\Delta\mu}{\varpi\,\left(-i\,\omega+\omega_{gap}\right)}\,.
\end{align}
In the above we have set,
\begin{align}
\vartheta=\frac{8\,\pi}{s_c\,q^2\,\rho^{(0)2}}\,\left(\varrho^{(0)}-\varrho \right)\,.
\end{align}

An obvious remaining question regards turning on a source $\delta \tilde{s}_Y$ for the operator $\mathcal{O}_Y$. As explained in \cite{Donos:2021pkk}, in the spontaneous case this is achieved by considering a non-trivial source term $\delta\theta_{(s)}$ in the UV expansion \eqref{eq:uv_bexp} for the bulk vector field. However, as shown in \cite{Donos:2021pkk}, turning a source $\lambda$ for the complex operator $\mathcal{O}_\psi$ would imply the non-conservation of electric charge through the Ward identity, 
\begin{align}\label{eq:current_non_cons}
\nabla_\alpha\langle J^\alpha\rangle=iq\,\left(\langle \mathcal{O}_\psi\rangle \lambda^\ast-\langle \mathcal{O}_\psi^\ast\rangle \lambda\right)\,.
\end{align}
Applying this for the case of our perturbative setup gives,
\begin{align}\label{eq:current_non_consV2}
\nabla_\alpha\langle J^\alpha\rangle=q\,\langle\mathcal{O}_\rho\rangle_b\,\delta \tilde{s}_{Y}\,.
\end{align}
Having it mind that the frequency is of order $\mathcal{O}(\varepsilon_\ast^2)$ and the leading correction to the charge density is of order $\mathcal{O}(\varepsilon_\ast)$, we see that the source $\delta\tilde{s}_Y$ should be of order $\mathcal{O}(\varepsilon_\ast^2)$ and we will write $\delta\tilde{s}_Y=\frac{\varepsilon_\ast^2}{2}\,\delta \tilde{s}_{Y (2)}+\cdots$.  Combining the above ingredients leads us to the charge imbalance,
\begin{align}\label{eq:charge_imbalance}
-i\omega_{[2]}\,\left(\delta\varrho_{\ast(2)}-\delta\varrho_{\#(2)} \right)=\frac{q}{2}\,\delta\langle\mathcal{O}_\rho\rangle_{\ast(0)}\,\delta\tilde{s}_{Y(2)}\,,
\end{align}
when linearly superposing the static perturbations obtained from the broken and the normal phase black holes.

We see that this source implies a slight modification of our previous derivations to account for the non-zero difference between the charge densities. One of the crucial steps concerns equation \eqref{eq:bulk_phase_sol} which now gets modified to,
\begin{align}\label{eq:bulk_phase_solV2}
\delta\tilde{\theta}_{(0)}=\frac{4\pi}{s_c\,q^2\,\delta\rho_{\ast(0)}^{(0)\,2}}\,\left(\delta\varrho^{(0)}_{\ast(2)}-\delta\varrho_{\ast(2)}+\frac{i\,q\,\delta\langle\mathcal{O}_\rho\rangle_{\ast(0)}}{2\,\omega_{[2]}}\,\delta\tilde{s}_{Y(2)}\right)\,.
\end{align}
Given the conditions \eqref{eq:current_non_consV2}, \eqref{eq:charge_imbalance} and \eqref{eq:bulk_phase_solV2}, evaluated the radial component of the symplectic current \eqref{eq:pr2} yields,
\begin{align}\label{eq:sY_source}
\delta\tilde{s}_Y=-\frac{\varpi\,\left(-i\omega+\omega_{gap} \right)}{\varepsilon_\ast\,\Delta\langle \mathcal{O}_\rho\rangle}\,\frac{\omega}{\omega\,\vartheta+2\,i\,\Delta\mu}\,.
\end{align}

We are now in the position to evaluate the retarded Green's functions,
\begin{align}
G_{\mathcal{O}_Y \mathcal{O}_Y}(\omega)&=\frac{q\,\langle\mathcal{O}_\rho\rangle\,\delta\tilde{\theta}_{(v)}}{\delta\tilde{s}_Y}\nn
&=-\frac{q^2\,\Delta\langle\mathcal{O}_\rho\rangle^2\,\left(\omega\,\vartheta+2\,i\,\Delta\mu\right)^2}{\omega^2\,\varpi\,\left(-i\,\omega+\omega_{gap}\right)}+\frac{i\,q^2\,\gamma\,\Delta\langle\mathcal{O}_\rho\rangle^2}{\omega}-\frac{2\,q^2\,\Delta\langle\mathcal{O}_\rho\rangle^2}{\omega^2}\,\left. \frac{\partial\mu_\#}{\partial\varrho}\right|_{s,\phi_{(s)}}\,,\nn
G_{\mathcal{O}_\phi \mathcal{O}_Y}(\omega)&=\frac{\delta\tilde{\langle\mathcal{O}_\phi\rangle}}{\delta\tilde{s}_Y}=\frac{2\,\Delta\langle\mathcal{O}_\phi\rangle}{\varepsilon_\ast\,\delta\tilde{s}_Y}+\frac{2\,i\,q\,\Delta\langle\mathcal{O}_\rho\rangle}{\omega}\,\left. \frac{\partial\langle\mathcal{O}_\phi\rangle_\#}{\partial \rho}\right|_{s,\phi_{(s)}}\nn
&=-\frac{2\,q\,\Delta\langle\mathcal{O}_\rho\rangle\,\Delta\langle\mathcal{O}_\phi\rangle}{\omega}\,\frac{\omega\,\vartheta+2\,i\,\Delta\mu}{\varpi\,\left(-i\,\omega+\omega_{gap}\right)}+\frac{2\,i\,q\,\Delta\langle\mathcal{O}_\rho\rangle}{\omega}\,\left. \frac{\partial\langle\mathcal{O}_\phi\rangle_\#}{\partial \rho}\right|_{s,\phi_{(s)}}\,,\nn
G_{\mathcal{O}_\rho \mathcal{O}_Y}(\omega)&=\frac{\delta\tilde{\langle\mathcal{O}_\rho\rangle}}{\delta\tilde{s}_Y}=-\frac{q\,\Delta\langle\mathcal{O}_\rho\rangle^2}{\omega}\,\frac{\omega\,\vartheta+2\,i\,\Delta\mu}{\varpi\,\left(-i\,\omega+\omega_{gap}\right)}\,,
\end{align}
after fixing,
\begin{align}
\gamma=\frac{4\,\pi}{s_c\,q^2\,\rho^{(0)2}}=\frac{1}{\varpi}\left(\vartheta^2+\frac{1}{q^2} \right)\,.
\end{align}
The last equality can be used to show that our expression for $G_{\mathcal{O}_Y \mathcal{O}_Y}$ leads to positive spectral weight. Moreover, the double pole in the same Green's function is due to the presence of the Goldstone mode, directly related to the phase of the condensate.

After noting that $\mathcal{O}_Y$ transforms as a pseudoscalar operator under time reversal, our expressions for the retarded Green's functions satisfy the Onsager relations\footnote{In order to check this, it is useful to note the Maxwell type of relation $\left.\frac{\partial\langle\mathcal{O}_\phi\rangle}{\partial\varrho}\right|_{s,\phi_{(s)}}=-\left.\frac{\partial\mu}{\partial\phi_{(s)}}\right|_{s,\varrho}$ coming from the enlarged first law \eqref{eq:first_law}}. Finally, our Green's function $G_{\mathcal{O}_Y \mathcal{O}_Y}$ agrees\footnote{In order to see the agreement, one has to write the transport coefficient $\Xi$ of \cite{Donos:2021pkk} as $\Xi\approx\gamma\,\chi_{QQ}^2/2$ in the near critical point limit. Moreover, at zero chemical potential we have that $\frac{\partial\mu_\ast}{\partial\varrho}=\chi_{QQ}^{-1}$. Note also that there is a mismatch of a factor of $2$ due to the different normalisation of the kinetic term of the complex scalar in our bulk action \eqref{eq:bulk_action_charged}.} with our previous result in \cite{Donos:2021pkk} at zero chemical potential and when we take the near critical point and zero pinning limit of the result presented in \cite{Donos:2021pkk}.

\section{Numerical checks}\label{sec:numerics}

The aim of this section is to provide numerical evidence in support of the analytic expression for the gapped mode given in equation \eqref{eq:Higgs_gap_neutral} and \eqref{eq:gap_super}. We consider the model  \eqref{eq:bulk_action} and \eqref{eq:bulk_action_charged_V2} respectively,
together with,
\begin{align} 
&V=-6+\frac{m_\rho^2}{2} \rho^2+ \frac{\lambda_\rho}{2}\, \rho^4+\frac{m_\phi^2}{2}\phi^2+\frac{\lambda_\phi}{2}\phi^4+ \lambda \,\rho^2\phi^2\,\nonumber\\
&\tau=1+\zeta_\rho \rho^2+\zeta_\phi \phi^2
\end{align}
and $m_\rho^2=-2\,,m^2_\phi=-2$.  As described in section \ref{sec:setup}, we choose to deform our boundary theory by a chemical potential $\mu=1$ and a relevant operator, $O_\phi$, with scaling dimension $\Delta_\phi= 2$; in particular we pick the source for $\phi$ to be given by $\phi_s=1$. The corresponding backreacted solution will then be given by black holes supported by a non-trivial profile for the scalar field $\phi$, while the scalar field $\rho$ remains trivial. These solutions correspond to the normal phase of the system.

\subsection{Ungauged model}
For the ungauged model with $\zeta_\rho=1, \zeta_\phi=0, \lambda_\phi=1,\lambda_\rho=1, \lambda=-1$, the normal phase black holes exhibit a second order phase transition at temperature  $T_c=0.115$ to a configuration supported by a non-trivial scalar field $\rho$. The latter describe the broken phase that we are interested in.

With these broken phase black holes at hand, we now turn our attention to studying perturbations around them. In order to check numerically the validity of the analytic expressions for the gap, we need to construct the quasinormal modes. In particular, we consider the following perturbations,
\begin{align}
&\delta g_{\mu\nu}= e^{-i \omega\, v(t,r)}\left(-U \delta g_{tt}(r)+e^{2 g(r)} \delta g_{xx}(r) (dx^2+dy^2)\right)\nonumber\\
&\delta A=e^{-i \omega\, v(t,r)}\delta a_t(r)\, dt\,,\nn
&\delta \phi=e^{-i \omega\, v(t,r)}\delta \phi(r)\,,\nn
&\delta \rho=e^{-i \omega\, v(t,r)}\delta \rho(r)\,,
\end{align}
where $v$ is the infalling Eddington-Finkelstein coordinate defined as,
\begin{equation}
v(t,r)=t+\int_\infty^r\frac{dy}{U(y)}\,.
\end{equation}
These perturbations satisfy three first order and two second order differential equations, which we solve using a shooting method subject to appropriate boundary conditions. In particular, in the horizon (located at r=0) we impose,
\begin{align}
&\delta g_{tt}=c_1\, r+\dots\,,\nn
&\delta g_{xx}=c_2 \,r+\dots\,,\nn
&\delta a_t=c_3\, r+\dots\,,\nn
&\delta \rho=\delta\rho_h+\dots\,,\nn
&\delta \phi=\delta\phi_h+\dots\,,
\end{align}
where $c_1, c_2, c_3$ are fixed, while close to the boundary we impose,
\begin{align}\label{eq:pertUV}
&\delta g_{tt}=\delta g_{tt}^{(s)}+\dots\,,\nn
&\delta g_{xx}=\delta g_{xx}^{(s)}+\dots\,,\nn
&\delta a_t=\delta a^{(s)}+\dots\,,\nn
&\delta \rho=\frac{\delta\rho_s}{r}+\frac{\delta\rho_v}{r^2}+\dots\,,\nn
&\delta \phi=\frac{\delta\phi_s}{r}+\frac{\delta\phi_v}{r^2}+\dots\,,
\end{align}
For the computation of quasinormal modes, we need to ensure that we remove all the sources from the UV expansion up to a combination of coordinate reparametrisations and gauge transformations, 
\begin{align}\label{eq:coord_transf}
[\delta g_{\mu\nu}+\mathcal{L}_\zeta g_{\mu\nu}]\to 0\,,\notag\\
[\delta A+\mathcal{L}_\zeta A+d\Lambda]\to 0\,,\notag\\
[\delta \phi+\mathcal{L}_\zeta \phi]\to 0\,,\notag\\
[\delta \rho+\mathcal{L}_\zeta \rho]\to 0\,.
\end{align} 
where the gauge transformations are of the form,
\begin{align}\label{eq:coord_transf}
x^\mu\to x^\mu+\zeta^\mu\,\quad\zeta&=e^{-i\omega t}\,\zeta^\mu \,\partial_\mu\,,\notag\\
A_\mu\to A_\mu+\partial_\mu \Lambda\,\quad\Lambda&=e^{-i\omega t}\,\lambda\,,
\end{align}
for $\zeta$, $\lambda$ constants. This requirement boils down to the sources appearing in \eqref{eq:pertUV} taking the form,
 \begin{align}
& \delta g_{t t}^{(s)}=2 i \omega\, \zeta_1-2 \zeta_2 \,,\notag\\
&\delta g_{x x}^{(s)}=-2\,\zeta_2\,,\notag\\
&\delta a^{(s)}=i \omega ( \mu\,\zeta_1+\lambda)\,,\notag\\
&\delta\rho_s=0\,,\notag\\
&\delta\phi_s=\phi_s\,\zeta_2\,.
 \end{align}

\begin{figure}[h!]
\centering
\includegraphics[width=0.8\linewidth]{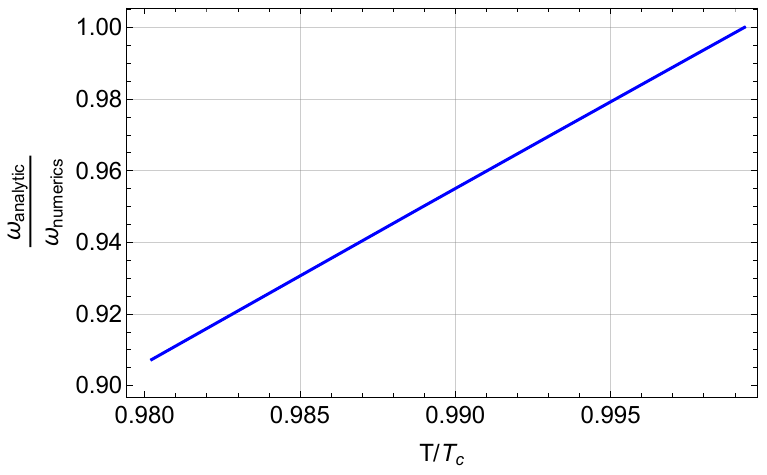}
\caption{Plot of the ratio of the analytic result for the gap coming from \ref{eq:Higgs_gap_neutral} and the gap computed numerically through the calculation of quasinormal modes as a function of  $T/T_c$. We see good quantitative agreement close to the critical temperature. Here $\phi_s=1,\mu=1,\zeta_\rho=1, \zeta_\phi=0,\lambda_\rho=1, \lambda_\phi=1$ and  $\lambda=-1$.}
\label{fig:tst}
\end{figure}

Overall, in addition to the frequency $\omega$, we have two constants in the IR ($\delta\rho_h,\delta\phi_h$) and five constants in the UV ($\zeta_1,\zeta_2,\lambda,\delta \rho^{(v)},\delta \phi^{(v)}$). This gives a total of eight free constants, out of which one can be set to unity due to the linearity of the equations. The remaining seven parameters match exactly the integration constants of the problem, giving rise to a single (discrete set) solution. In figure \ref{fig:tst}, we compare the results for gap coming from the direct calculation of the quasinormal modes  and from the analytic expression \eqref{eq:Higgs_gap_neutral} for $\zeta_\rho=1, \zeta_\phi=0, \lambda_\phi=1,\lambda_\rho=1$, and  $\lambda=-1$. In particular, we plot the ratio $\omega_{analytic}/\omega_{numerics}$ as a function of $T/T_c$. We see good quantitative agreement as $T\to T_c$.

\subsection{Superfluid model}

For superfluids with $\zeta_\rho=1, \zeta_\phi=0, \lambda_\phi=1,\lambda_\rho=1, \lambda=-1$, the critical temperature is  $T_c=0.128$ and it also corresponds to a second order phase transition. The broken phase is supported by a non-trivial charged condensate, $\rho$. Given these backreacted black holes, we now turn our attention to studying perturbations around them. 

\begin{figure}[h!]
\centering
\includegraphics[width=0.8\linewidth]{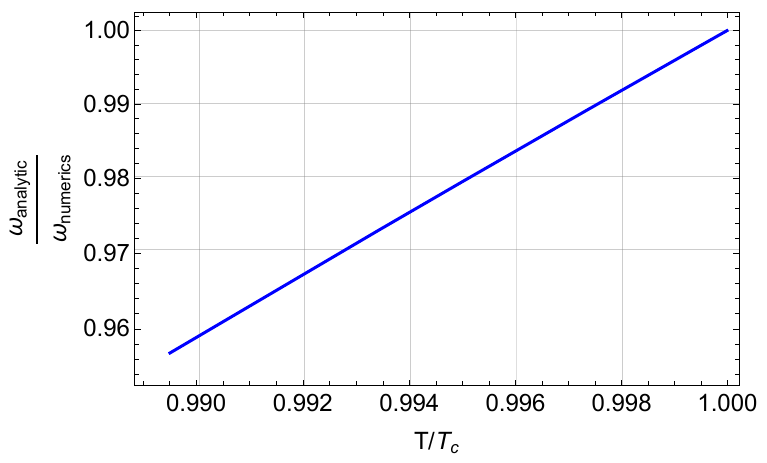}
\caption{Comparison of the analytic and numerical result for the gap for superfluids. Here $\phi_s=1, q=1,\mu=1,\zeta_\rho=1, \zeta_\phi=0,\lambda_\rho=1, \lambda_\phi=1$ and  $\lambda=-1$.}
\label{fig:tst2}
\end{figure}

Similarly to above, to check the validity of the analytic expressions for the gap, we construct the quasinormal modes. In particular, we consider the following perturbations,
\begin{align}
&\delta g_{\mu\nu}= e^{-i \omega\, v(t,r)}\left(-U \delta g_{tt}(r)+e^{2 g(r)} \delta g_{xx}(r) (dx^2+dy^2)\right)\nonumber\\
&\delta B=e^{-i \omega\, v(t,r)} (\delta b_t(r)\, dt+\delta b_r(r) dr)\,,\nn
&\delta \phi=e^{-i \omega\, v(t,r)}\delta \phi(r)\,,\nn
&\delta \rho=e^{-i \omega\, v(t,r)}\delta \rho(r)\,,
\end{align}
where $v$ is the infalling Eddington-Finkelstein coordinate. These perturbations satisfy three first order and three second order differential equations, which we solve using a shooting method subject to the following boundary conditions. Close to the horizon (located at r=0) we impose,
\begin{align}
&\delta g_{tt}=c_1\, r+\dots\,,\nn
&\delta g_{xx}=c_2 \,r+\dots\,,\nn
&\delta b_t=\delta b_t^{h}+\dots\,,\nn
&\delta b_r=\frac{\delta b_t^{h}}{4 \pi \,T\, r}+\dots\,,\nn
&\delta \rho=\delta\rho^h+\dots\,,\nn
&\delta \phi=\delta\phi^h+\dots\,,
\end{align}
where $c_1, c_2$ are fixed, while close to the boundary we require that,
\begin{align}\label{eq:pertUV_super}
&\delta g_{tt}=\delta g_{tt}^{(s)}+\dots\,,\nn
&\delta g_{xx}=\delta g_{xx}^{(s)}+\dots\,,\nn
&\delta b_t=\delta b_t^{(s)}+\dots\,,\nn
&\delta b_r=0+\dots\,,\nn
&\delta \rho=\frac{\delta\rho_s}{r}+\frac{\delta\rho_v}{r^2}+\dots\,,\nn
&\delta \phi=\frac{\delta\phi_s}{r}+\frac{\delta\phi_v}{r^2}+\dots\,,
\end{align}
Demanding the absence of sources, the sources appearing in \eqref{eq:pertUV_super} take the form,
 \begin{align}
& \delta g_{t t}^{(s)}=2 i \omega\, \zeta_1-2 \zeta_2 \,,\notag\\
&\delta g_{x x}^{(s)}=-2\,\zeta_2\,,\notag\\
&\delta b_t^{(s)}=i \omega ( \mu\,\zeta_1+\lambda-\delta \theta_{(v)})\,,\notag\\
&\delta\rho_s=0\,,\notag\\
&\delta\phi_s=\phi_s\,\zeta_2\,,
 \end{align}
Overall, in addition to the frequency $\omega$, we have three constants in the IR ($\delta\rho^h,\delta\phi^h,\delta b_t^{h}$) and six constants in the UV ($\zeta_1,\zeta_2,\lambda,\delta \rho^{(v)},\delta \phi^{(v)}, \delta \theta_{(v)}$). Due to the linearity of the equations, one out of these ten free constants can be set to unit. One is then left with nine parameters which match exactly the integration constants of the problem, giving rise to a single (discrete set) solution. Just like above, in figure \ref{fig:tst2}, we plot the ratio $\omega_{analytic}/\omega_{numerics}$ coming from the analytic expression \eqref{eq:gap_super} and from the direct calculation of the quasinormal modes for  $\zeta_\rho=1, \zeta_\phi=0, \lambda_\phi=1,\lambda_\rho=1$, and  $\lambda=-1$. We see good quantitative agreement as $T\to T_c$.

For this model with $\zeta_\rho=0, \zeta_\phi=0,  \lambda_\phi=1, \lambda_\rho=1/4, \lambda=-3/4,$ and $T_c=0.0484121$, we also compute the Green's functions for the scalar fields for $T/T_c=0.999957$, corresponding to $\omega_{gap}=4\cdot 10^{-5}$. In this case, the IR and UV expansions stay unchanged. For the metric and gauge field sources we impose absence of sources just like above. For the scalar field sources, we impose,
 \begin{align}
&\delta\rho_s=s_1\,,\notag\\
&\delta\phi_s=s_2+\phi_s\,\zeta_2\,,
 \end{align}
and we set either $(s_1,s_2)=(1,0)$ or $(s_1,s_2)=(0,1)$ depending on which correlator we want to compute.  Note that under the reparametrisation \eqref{eq:coord_transf}, the VEVs for the scalars transform as
 \begin{align}
&\delta\rho_v\to \delta\rho_v-2 \rho_v\,\zeta_2 \,,\notag\\
&\delta\phi_v\to \delta\phi_v-2 \phi_v\,\zeta_2 +i \omega\, \phi_s\,\zeta_2\,,
 \end{align}

\begin{figure}[h!]
\centering
\includegraphics[width=0.48\linewidth]{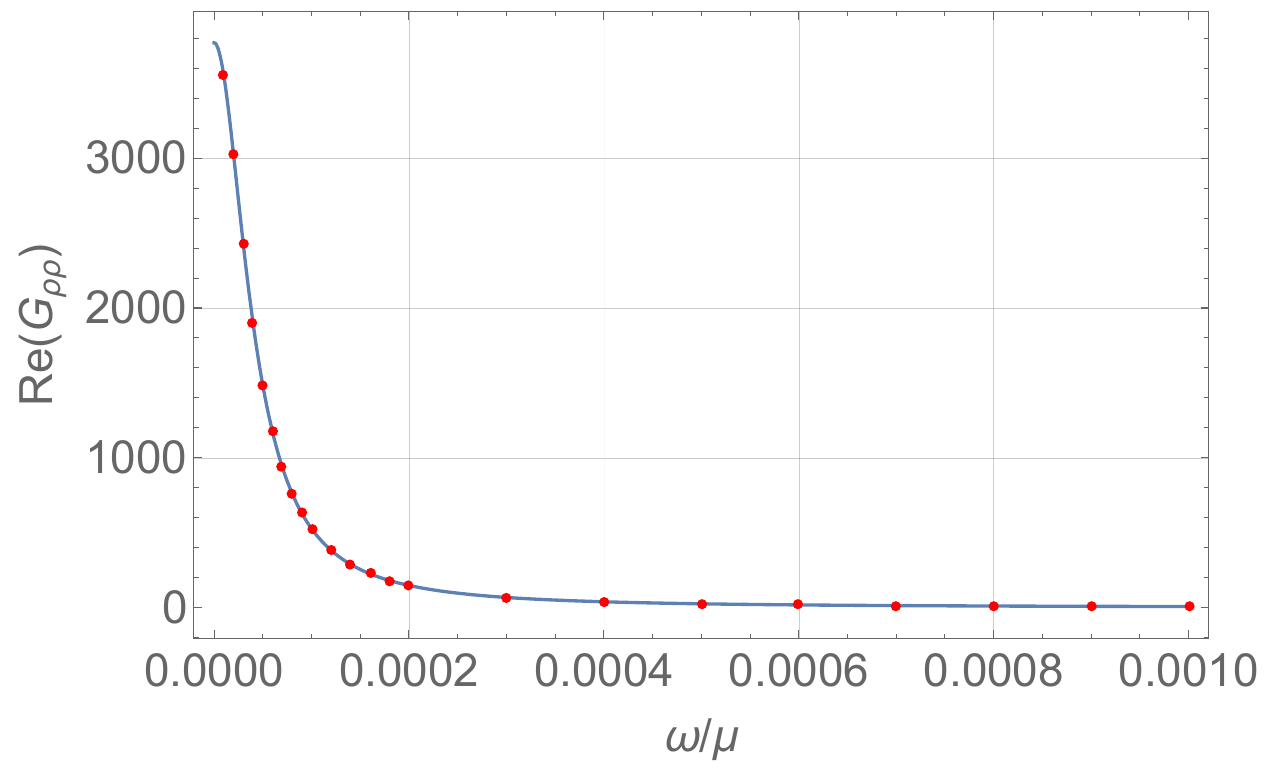}\quad \includegraphics[width=0.48\linewidth]{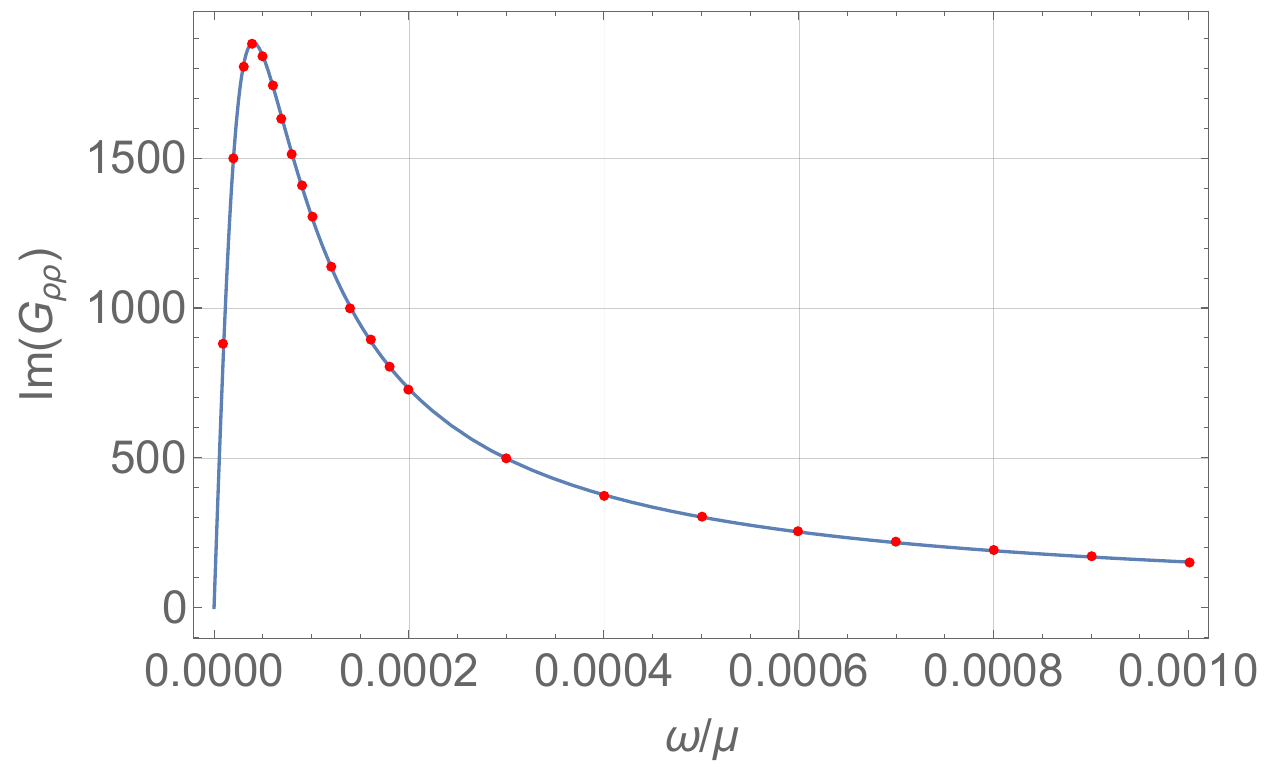}\\
\includegraphics[width=0.48\linewidth]{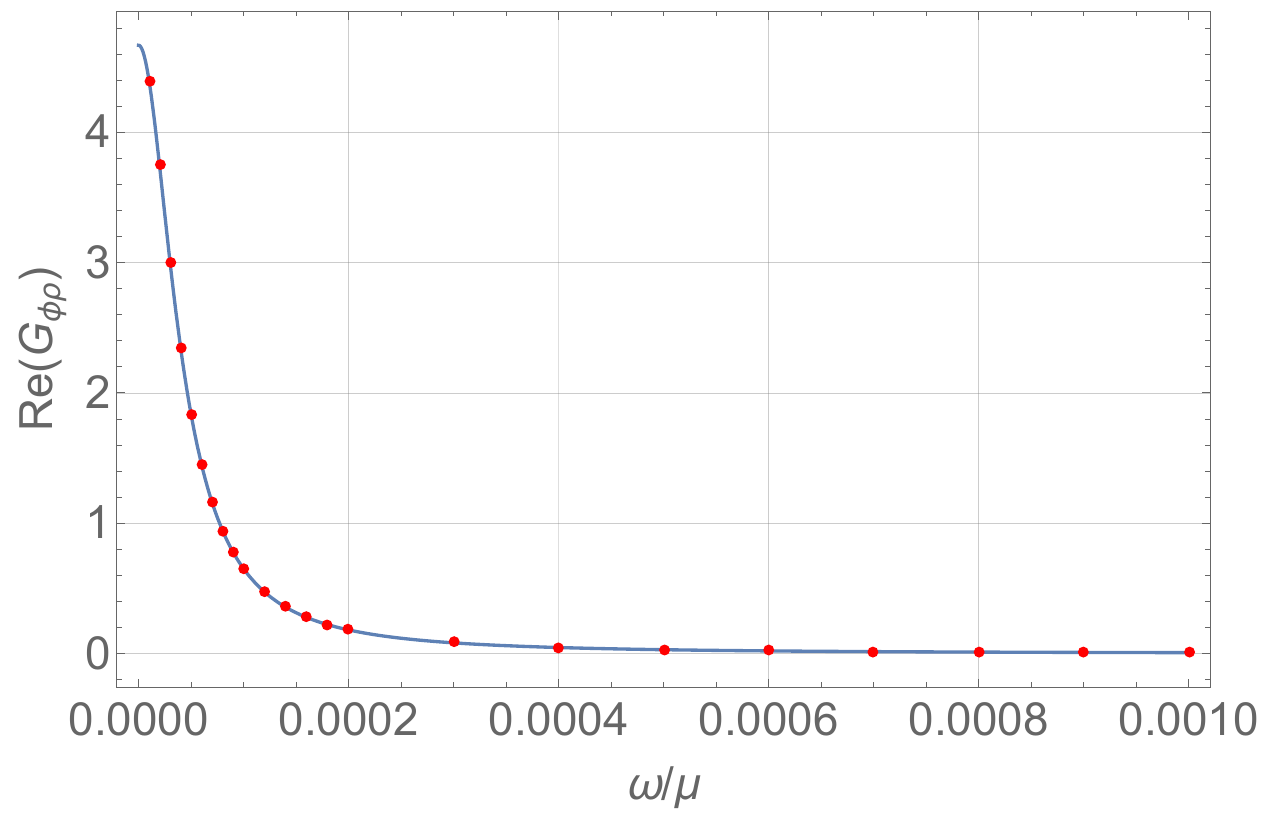}\quad \includegraphics[width=0.48\linewidth]{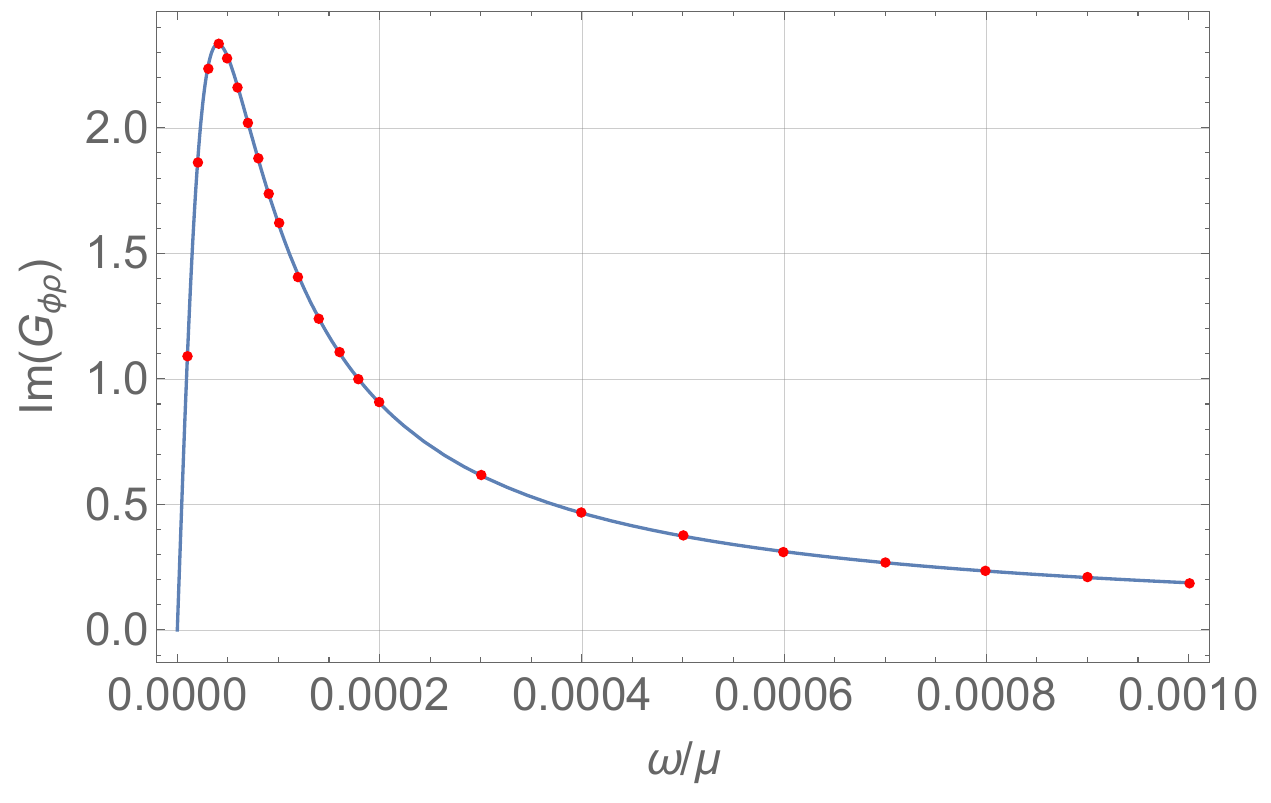}
\caption{Real and Imaginary part of the Green's functions. Analytic results are displayed in dashed lines, while numerical results are show with black dots. Here $\phi_s=1, q=1, \mu=1,\zeta_\rho=0, \zeta_\phi=0, \lambda_\phi=1, \lambda_\rho=1/4, \lambda=-3/4$ and $T/T_c=0.999957$.}
\label{fig:tst3}
\end{figure}

\begin{figure}[h!]
\centering
\includegraphics[width=0.48\linewidth]{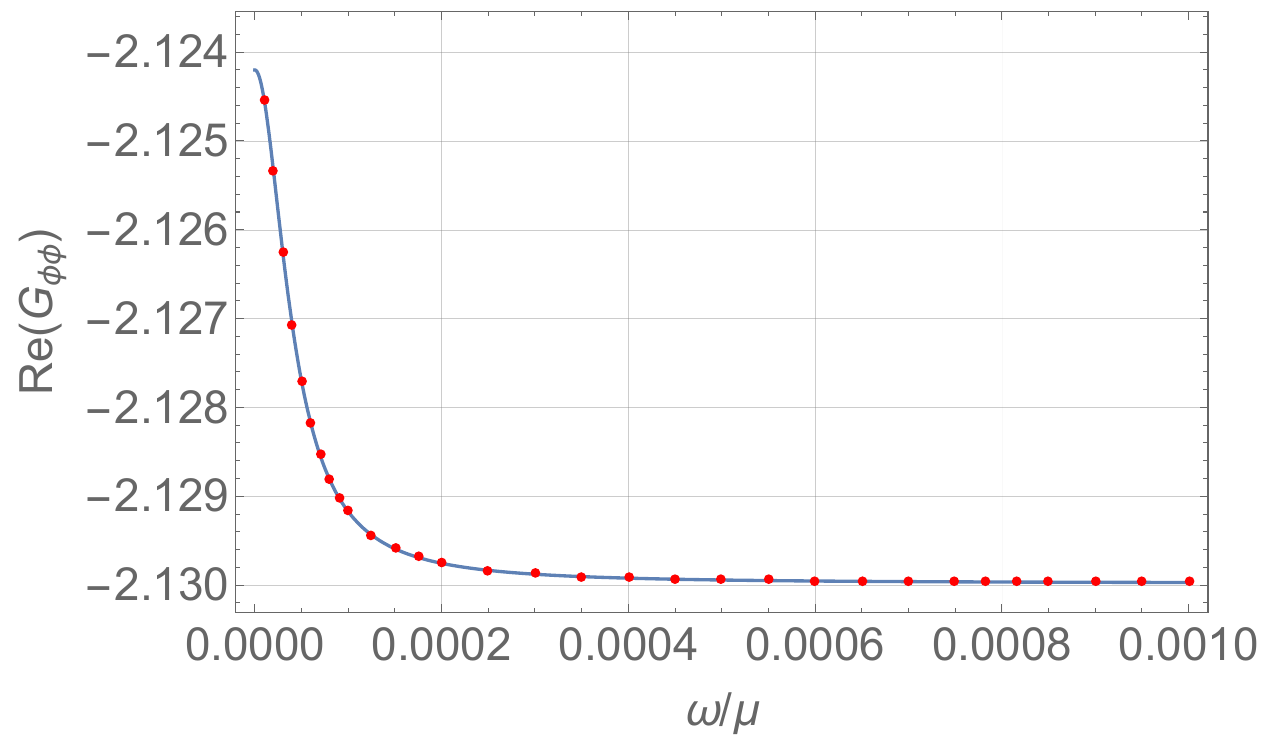}\quad \includegraphics[width=0.48\linewidth]{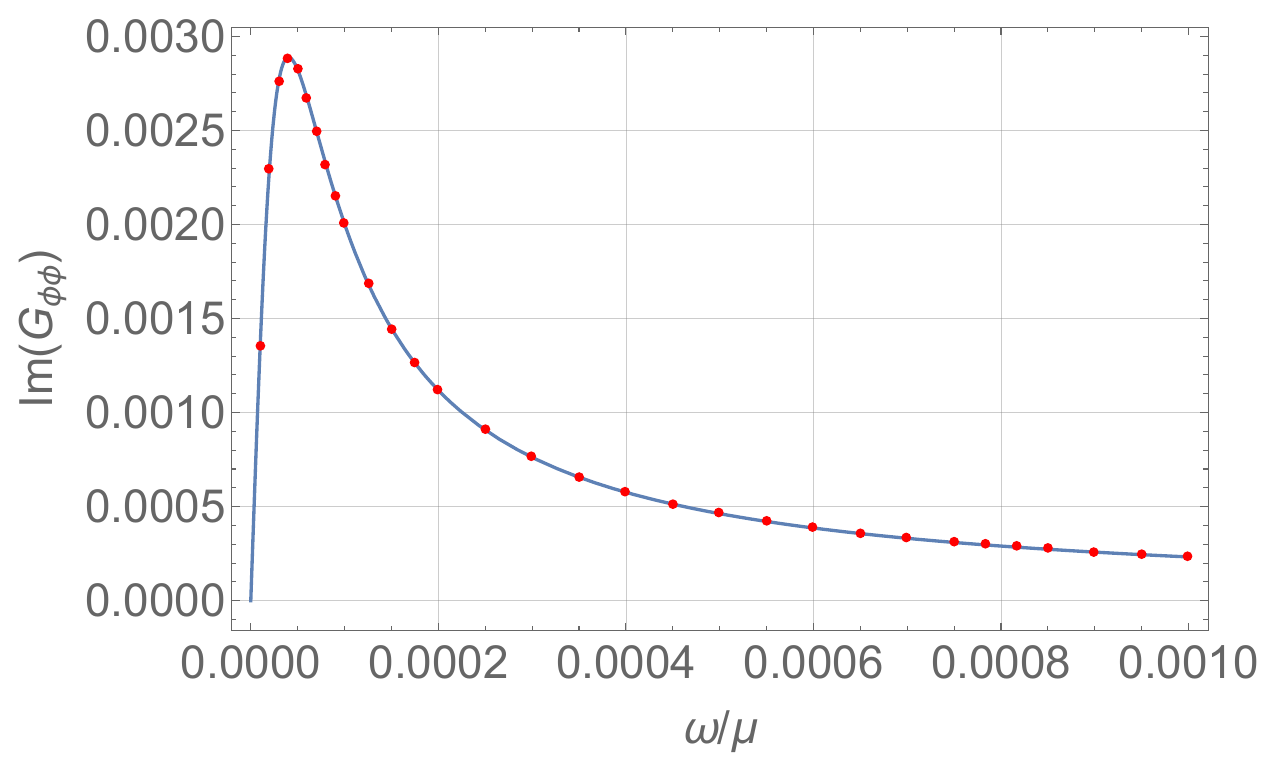}\\
\includegraphics[width=0.48\linewidth]{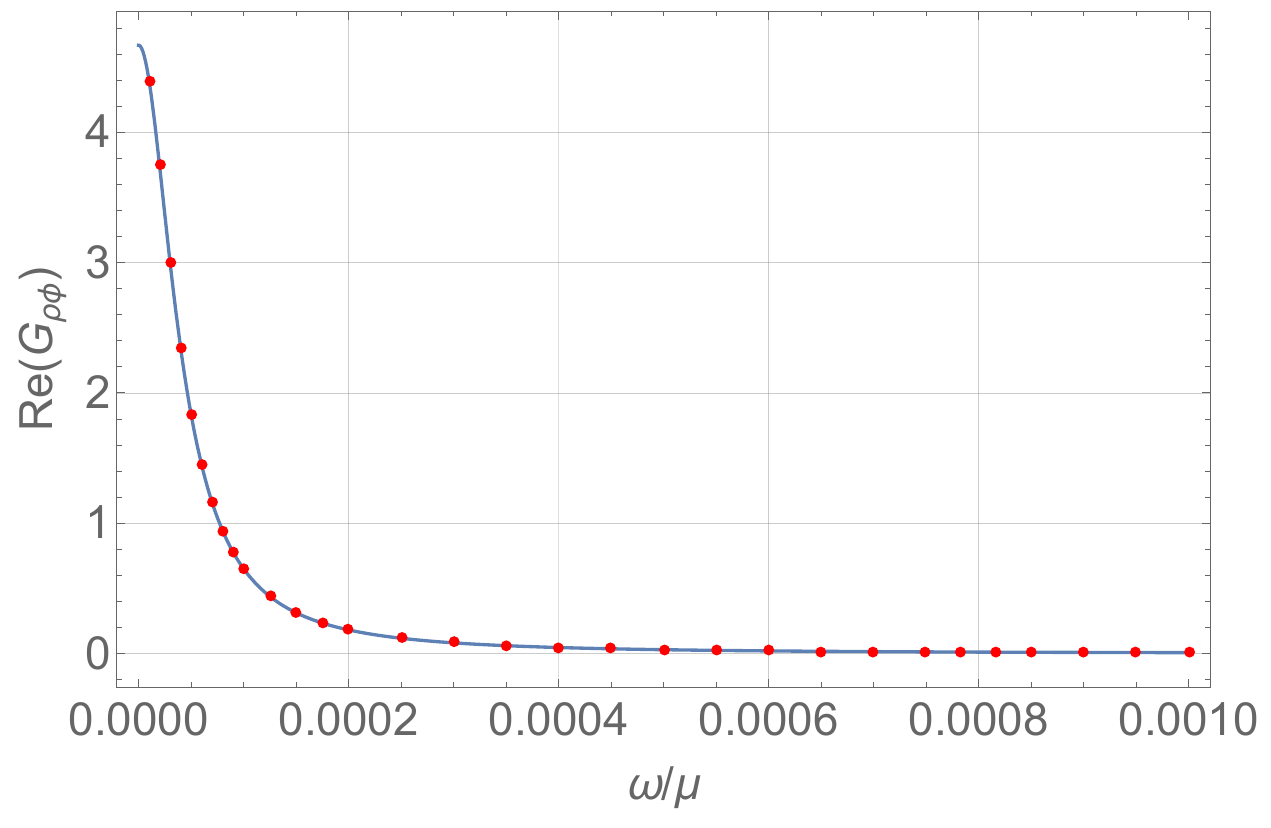}\quad \includegraphics[width=0.48\linewidth]{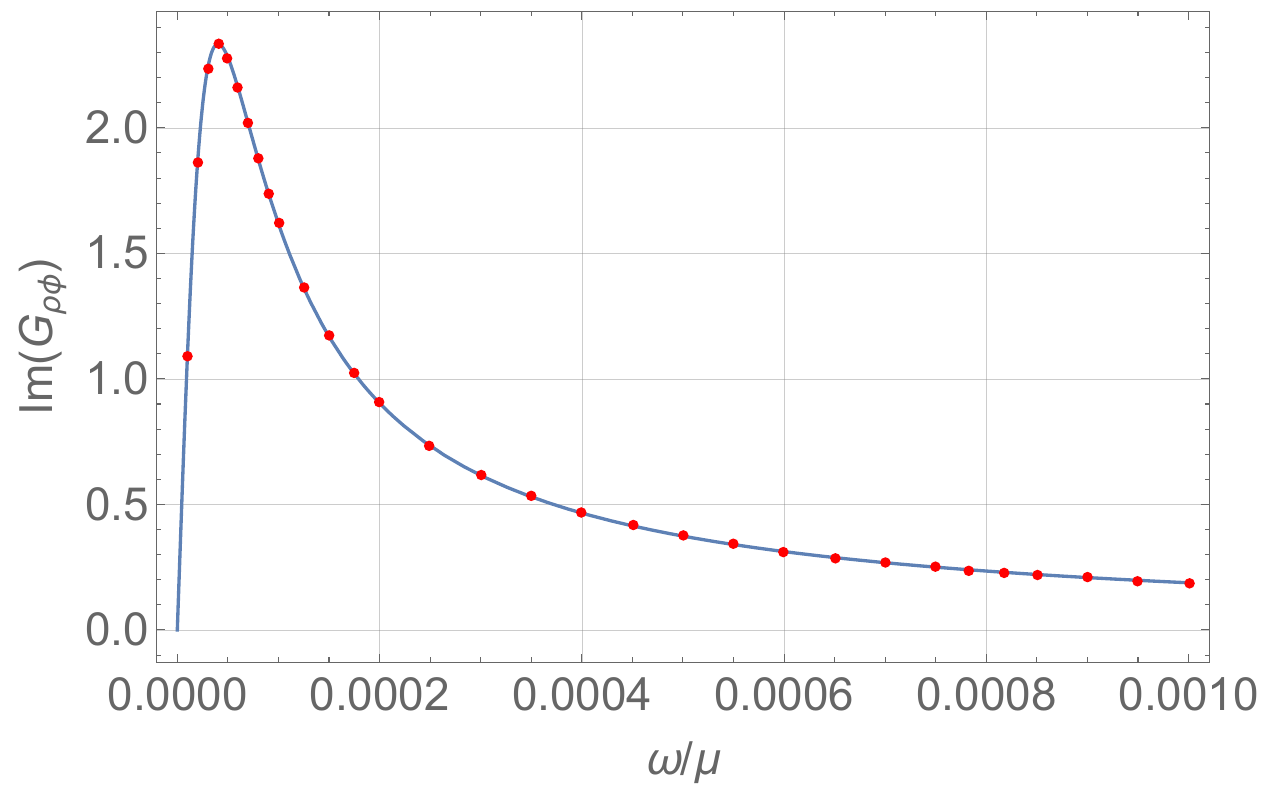}
\caption{Real and Imaginary part of the Green's functions. Analytic results are displayed in dashed lines, while numerical results are show with black dots. Here $\phi_s=1, q=1, \mu=1,\zeta_\rho=0, \zeta_\phi=0, \lambda_\phi=1, \lambda_\rho=1/4, \lambda=-3/4$ and $T/T_c=0.999957$.}
\label{fig:tst4}
\end{figure}

Overall, for fixed $s_1,s_2$, we have nine constants in the expansion $\delta\rho^h,\delta\phi^h,\delta b_t^{h}, \zeta_1$, $\zeta_2,\lambda,\delta \rho^{(v)},\delta \phi^{(v)}, \delta \theta_{(v)}$ in addition to the frequency $\omega$. Given that the problem is fixed in terms of nine integration constants, we conclude that we expect to find a one parameter family of solutions labeled by $\omega$. In figure \ref{fig:tst3} and \ref{fig:tst4}, we plot the real and imaginary parts of the four Green's functions in terms of the frequency: the red dots correspond to the numerical results, while the solid blue lines indicate the analytic expressions of the previous section. We see excellent quantitative agreement. Note that the agreement is expected to improve as one approaches the critical temperature. The location of the peak in the
imaginary part of the Green’s functions corresponds to the location of the
gap.


\newpage

\section{Discussion}\label{sec:discussion}

We considered the perturbative dynamics of holographic systems which are parametrically close to a thermal phase transition. In particular, we focused on gravitational fluctuations capturing the Higgs mode which universally emerges in the broken phase of all second order phase transitions.

One of the crucial ingredients in our construction is the utilisation of the Crnkovic-Witten symplectic current. In the case of the source free quasinormal mode, dual to the Higgs mode, this allowed us to obtain the gap of equations \eqref{eq:Higgs_gap_neutral} and \eqref{eq:gap_super}. In the case of Section \ref{sec:greens}, where the frequency was fixed by an external source, the Crnkovic-Witten symplectic current allowed us to relate the external source to the VEVs of our scalar operators, fixing the form of the Green's functions. The second important ingredient was the construction of the terms which are next to leading order in the $\varepsilon$ expansion of equation \eqref{eq:timed_pert}. As we explained in section \ref{sec:t_pert}, these are simply linear combinations of the static perturbations that can be obtained by varying the thermodynamic parameters of the broken and normal phase black holes.

The results of Section \ref{sec:greens} confirm that the Higgs/amplitude mode dominates the linear response of the scalar operators in the infinite wavelength limit. At finite wavelengths, the Higgs mode will mix with the charge and current densities of theory, affecting the hydrodynamics. It is natural to consider holographic techniques in order to derive an enlarged hydrodynamic description which includes the dynamics of the Higgs mode. This, would increase the range of validity of hydrodynamics close to the critical point.

A different physical situation in which a parametrically small gap exists is when a symmetry is broken pseudospontanesouly. This is the case when apart from the spontaneous breaking, a small explicit parameter that breaks the symmetry is introduced in the system. Our recent work \cite{Donos:2021pkk} makes clear that our techniques are applicable in the case of superfluids with the internal broken pseudosponaneously. It would be interesting to apply our techniques in setups where the weakly broken symmetry is translations \cite{Andrade:2017cnc,Amoretti:2019kuf,Baggioli:2021xuv}.

In this paper, we have considered phase transitions which preserve the spacetime symmetries of the boundary theory. However, over the past few years a plethora of gravitational instabilities which spontaneously break translations have been discovered in the context of holography \cite{Nakamura:2009tf,Donos:2011bh,Donos:2011ff,Donos:2011qt}. These have been shown to lead to second order phase transitions in general \cite{Donos:2012wi,Donos:2013wia,Withers:2013loa,Withers:2014sja,Donos:2015eew}. It is interesting to examine the dynamics of the Higgs/amplitude mode for the order parameter of density waves.

From a slightly different angle, in \cite{Donos:2021pkk} we used the Crnkovic-Witten current to obtain the dissipative terms in the constitutive relations for the electric current in the superfluid phase at zero charge density. Even though that setup is simple due to the non-mixing of the normal and the superfluid components, it can be seen as a proof of principle in extracting the transport coefficients of holographic theories.We expect that the techniques we have developed here and in \cite{Donos:2021pkk} will be central in the directions outlined above and we wish to report more on these in the near future.

Finally, it is necessary to make the connection between our results and the field theory approach on critical phenomena e.g. \cite{RevModPhys.49.435, Grossi:2021gqi, Rajagopal:1992qz}. In particular, it would be very interesting to understand the role of our transport coefficient $\varpi$ in the framework of field theory.

\section*{Acknowledgements}

We would like to thank P. Kailidis and V. Ziogas for discussions and collaboration in related topics. AD is supported by STFC grant ST/T000708/1. C.P. acknowledges support from a Royal Society - Science Foundation Ireland University Research Fellowship via grant URF/R1/211027.

\appendix


\section{Symplectic Current Contributions}\label{app:sympl_current_terms}
Here we list all the relevant terms that enter in the construction of the symplectic current \eqref{eq:scurrent_bulk}. After writing the bulk action \eqref{eq:bulk_action_charged_V2} in terms of first derivatives of the metric, we obtain,
\begin{align}\label{eq:sympl_cur_contr}
\frac{\partial\mathcal{L}}{\partial\partial_\mu\phi}&=-\sqrt{-g}\,\partial^\mu\phi,\quad \frac{\partial\mathcal{L}}{\partial\partial_\mu\rho}=-\sqrt{-g}\,\partial^\mu\rho,\quad
\frac{\partial\mathcal{L}}{\partial\partial_\mu A_\alpha}=-\sqrt{-g}\,Z\,F^{\mu\alpha}\,,\notag\\
\frac{\partial\mathcal{L}}{\partial\partial_\mu g_{\alpha\beta}}&=\sqrt{-g}\,\Gamma^\mu_{\gamma\delta}\left(\,g^{\gamma\alpha}\,g^{\delta\beta}-\frac{1}{2}\,g^{\gamma\delta}g^{\alpha\beta} \right)-\sqrt{-g}\,\Gamma^\kappa_{\kappa\lambda}\,\left(g^{\mu\left(\alpha\right.}g^{\left.\beta\right)\lambda}-\frac{1}{2}\,g^{\mu \lambda}g^{\alpha\beta}\right)\,,
\end{align}
where $\Gamma^{\alpha}_{\beta\gamma}$ are the Christoffel symbols compatible with our perturbed metric.

\section{The Energy Difference $\Delta E$}\label{app:thermo}
In this appendix we will show that equation \eqref{eq:energy_gap} gives the energy difference of the broken and the normal phase at fixed scalar deformation, entropy and charge density. The energy density is,
\begin{align}
E=F+T\,s+\mu\,\varrho,
\end{align}
satisfying the First Law of thermodynamics,
\begin{align}\label{eq:first_law}
\delta E=T\,\delta s+\mu\,\delta\varrho-\langle\mathcal{O}_\phi\rangle\,\delta\phi_{(s)}\,,
\end{align}
given that the gravitation free energy $F$ is the appropriate potential for the grand canonical ensemble,
\begin{align}
\delta F=-s\,\delta T-\varrho\,\delta\mu-\langle\mathcal{O}_\phi\rangle\,\delta\phi_{(s)}\,.
\end{align}
As we would expect the above suggests that energy should be considered as a function of $s$, $\varrho$ and $\phi_{(s)}$.

We could think of expanding the energy at the critical point along the broken and the normal phases. Instead of doing that separately, we will do this collectively for both the broken and the normal phase and specialise in the end, when we take the difference. Doing so we obtain,
\begin{align}\label{eq:energy_exp}
\delta E=T\,\delta s+\mu\,\delta\varrho-\langle\mathcal{O}_\phi\rangle\,\delta\phi_{(s)}+\frac{1}{2}\,\delta\Gamma^T
\begin{pmatrix}
\frac{\partial^2 E}{\partial s^2} & \frac{\partial^2 E}{\partial s \,\partial\varrho} & \frac{\partial^2 E}{\partial s\, \partial\phi_{(s)}}\\
\frac{\partial^2 E}{\partial s \,\partial\varrho} & \frac{\partial^2 E}{\partial \varrho^2} & \frac{\partial^2 E}{\partial\varrho\,\partial\phi_{(s)}} \\
\frac{\partial^2 E}{\partial s \,\partial\phi_{(s)}} & \frac{\partial^2 E}{\partial \varrho \,\partial\phi_{(s)}} & \frac{\partial^2 E}{\partial\phi_{(s)}^2}
\end{pmatrix}
\delta\Gamma
+\cdots\,,
\end{align}
where we have set,
\begin{align}
\delta\Gamma=
\begin{pmatrix}
\delta s \\ \delta\varrho \\ \delta\phi_{(s)}
\end{pmatrix}=
\begin{pmatrix}
\delta\mathbf{\Phi}\\ \delta\phi_{(s)}
\end{pmatrix}
\,.
\end{align}

Equation \eqref{eq:energy_exp} can be written as,
\begin{align}
\delta E=&T\,\delta s+\mu\,\delta\varrho-\langle\mathcal{O}_\phi\rangle\,\delta\phi_{(s)}\nn
&+\frac{1}{2}\,\delta\mathbf{\Phi}^T\,
\begin{pmatrix}
\frac{\partial^2 E}{\partial s^2} & \frac{\partial^2 E}{\partial s \,\partial\varrho}\\
\frac{\partial^2 E}{\partial s \,\partial\varrho} & \frac{\partial^2 E}{\partial \varrho^2} 
\end{pmatrix}\,
\delta\mathbf{\Phi}+\delta\mathbf{\Phi}^T\,
\begin{pmatrix}
\frac{\partial^2 E}{\partial s \,\partial\phi_{(s)}} \\ \frac{\partial^2 E}{\partial \varrho \,\partial\phi_{(s)}} 
\end{pmatrix}\,
\delta\phi_{(s)}+\frac{1}{2}\,\frac{\partial^2 E}{\partial\phi_{(s)}^2}\,\delta\phi_{(s)}^2+\cdots\nn
=&T\,\delta s+\mu\,\delta\varrho-\langle\mathcal{O}_\phi\rangle\,\delta\phi_{(s)}\nn
&+\frac{1}{2}\,\delta\mathbf{\Phi}^T\,
\begin{pmatrix}
\frac{\partial T}{\partial s} & \frac{\partial T}{\partial\varrho}\\
\frac{\partial \mu}{\partial s} & \frac{\partial \mu}{\partial \varrho} 
\end{pmatrix}\,
\delta\mathbf{\Phi}-\delta\mathbf{\Phi}^T\,
\begin{pmatrix}
\frac{\partial}{\partial s}\langle\mathcal{O}_\phi\rangle \\ \frac{\partial}{\partial \varrho }\langle\mathcal{O}_\phi\rangle 
\end{pmatrix}\,
\delta\phi_{(s)}-\frac{1}{2}\,\delta\phi_{(s)}^2\,\frac{\partial}{\partial\phi_{(s)}}\langle\mathcal{O}_\phi\rangle+\cdots\nn
=&T\,\delta s+\mu\,\delta\varrho-\langle\mathcal{O}_\phi\rangle\,\delta\phi_{(s)}\nn
&+\frac{1}{2}\,\delta\mathbf{\Phi}^T\,
\mathbf{\Xi}^{-1}\,
\delta\mathbf{\Phi}-\delta\mathbf{\Phi}^T\,
\mathbf{\Xi}^{-1}\,\mathbf{\nu}\,
\delta\phi_{(s)}-\frac{1}{2}\,\delta\phi_{(s)}^2\,\frac{\partial}{\partial\phi_{(s)}}\langle\mathcal{O}_\phi\rangle+\cdots\,,
\end{align}
where in the last line we recognised the matrix of susceptibilities as the Jacobian of the transformation when passing from $(s,\varrho)$ to $(T,\mu)$. For the last term we will consider the chain rule,
\begin{align}
\left.\frac{\partial}{\partial\phi_{(s)}}\langle\mathcal{O}_\phi\rangle\right|_{s,\varrho}&=\left.\frac{\partial}{\partial\phi_{(s)}}\langle\mathcal{O}_\phi\rangle\right|_{T,\mu}+\left.\frac{\partial T}{\partial\phi_{(s)}}\right|_{s,\varrho}\,\left.\frac{\partial}{\partial T}\langle\mathcal{O}_\phi\rangle\right|_{\mu,\phi_{(s)}}+\left.\frac{\partial \mu}{\partial\phi_{(s)}}\right|_{s\varrho}\,\left.\frac{\partial}{\partial \mu}\langle\mathcal{O}_\phi\rangle\right|_{T,\phi_{(s)}}\nn
&=\nu_\phi-\mathbf{\nu}^T\,\mathbf{\Xi}^{-1}\,\mathbf{\nu}\,.
\end{align}
where in the last line we used equation \eqref{eq:Tmu_relations2} to compute the partial derivatives of $T$ and $\mu$ at fixed $s$ and $\varrho$. Putting everything together we have that,
\begin{align}
\delta E=&T\,\delta s+\mu\,\delta\varrho-\langle\mathcal{O}_\phi\rangle\,\delta\phi_{(s)}\nn
&+\frac{1}{2}\,\delta\mathbf{\Phi}^T\,
\mathbf{\Xi}^{-1}\,
\delta\mathbf{\Phi}-\delta\mathbf{\Phi}^T\,
\mathbf{\Xi}^{-1}\,\mathbf{\nu}\,
\delta\phi_{(s)}-\frac{1}{2}\,\left(\nu_\phi-\mathbf{\nu}^T\,\mathbf{\Xi}^{-1}\,\mathbf{\nu}\right)\,\delta\phi_{(s)}^2+\cdots\,,
\end{align}
The above shows that after expanding along the broken and the normal phases and taking the difference, the terms linear in the variations will cancel out and the quadratic ones will give the result of equation \eqref{eq:energy_gap}.

\newpage
\bibliographystyle{utphys}
\bibliography{refs}{}
\end{document}